\documentclass[fleqn,10pt]{wlscirep}

\usepackage[T1]{fontenc}

\usepackage[utf8]{inputenc}
\usepackage{cite}

\usepackage{hyperref}
\usepackage{graphicx}
\usepackage{subcaption}
\usepackage{lipsum}
\usepackage{array}
\usepackage[utf8]{inputenc}
\usepackage{fontenc}
\DeclareUnicodeCharacter{2005}{ }

\title{Versatile High-Power Monolithic All-Glass Fiber Amplifier for Pulsed Signals with a Wide Range of Repetition Rates}

\author[1,*]{Hossein Fathi}
\author[2]{Ebrahim Aghayari}
\author[3]{Andrey Grishchenko}
\author[4]{Amit Yadav}
\author[4]{Edik Rafailov}
\author[1,2]{Regina Gumenyuk}
\author[2]{Valery Filippov}
\affil[1]{Laboratory of Photonics, Physics Unit, Faculty of Engineering and Natural Sciences, Tampere University, Korkeakoulunkatu 3, 33720 Tampere, Finland}
\affil[2]{Ampliconyx Ltd, Lautakatonkatu 18, 33580, Tampere, Finland}
\affil[3]{Ceramoptec SIA, Domes street 1a, Livani, LV-5316, Latvia}
\affil[4]{Aston Institute of Photonic Technologies (AIPT), Aston University, Birmingham, B4 7ET, U.K}
\affil[*]{hossein.fathi@tuni.fi}

\begin{abstract}

This study presents a compact, high-power monolithic all-glass spun tapered double-clad fiber amplifier for single-stage amplification of narrow linewidth picosecond pulsed signals from a few tens of mW to several hundred Watts of average power and MW level of peak power, covering a wide range of repetition rates. The absence of free-space elements in the amplifier module enhances its overall reliability by omitting the dependency on the pump alignment and internal back reflections. The versatile all-glass amplifier module delivers 50 ps pulses with over 2 MW peak power at 1 MHz, 50 ps pulses with over 625 W average power at 20 MHz, and 20 ps pulses with over 645 W average power at 1 GHz, all exhibiting excellent spectral, spatial, and polarization characteristics. This monolithic all-glass ultra-large mode area fiber amplifier is verified as a robust solution for direct amplification of short pulses attaining high peak/average power laser systems with excellent spectral, spatial, and polarization characteristics.
  
\end{abstract}
\begin{document}

\flushbottom
\maketitle
\thispagestyle{empty}

\section*{Introduction}

Over the past two decades, significant progress was made in fiber laser technology, spurred by its outstanding properties. As a result, fiber lasers experienced remarkable advancements, further solidifying their importance across a wide range of laser-based applications, from scientific research to industrial processes. They emerged as a strong alternative to traditional laser sources due to their numerous inherent benefits, including superior beam quality and stability, high quantum efficiency, broad gain bandwidth, excellent thermal management, and turn-key operation\cite{6808413}. Fiber lasers started slowly replacing traditional Ti:Sa lasers in the high-harmonic generation systems, offering higher repetition rates and low carrier-envelope phase noise. Providing sufficient average power scalability in combination with nonlinear post-compression, fiber-based laser systems are capable of keeping the required intensity level and transferring it to higher Extreme UV flux\cite{Raab:22, Shestaev:20, Shestaev:20-}. Furthermore, the amplification of high-repetition-rate (HR) pulses has gained significant interest in material processing for achieving both high-energy and high-peak-power pulses in kilowatt-class laser systems, particularly through burst-mode operation\cite{Burst-Liu:20, Burst-Liu_Zhang_Bu_Zhao_Zhu_Yang_Hou_2023, Burst-photonics11060570, Xiu:23}. This approach allows for the simultaneous realization of highly intense pulses while maintaining manageable thermal loads in the laser system.  By distributing energy over multiple bursts, the approach optimizes both power and performance, making it valuable for applications that require a balance between energy scaling and thermal management.

Amplifying short pulses directly within a single fiber laser system proved challenging, primarily due to thermally induced transverse mode instabilities and detrimental nonlinear effects.
The complexity scales up by a growing demand for compact and efficient short-pulsed amplifiers capable of delivering high-power/energy lasers with excellent beam profiles, pulse shapes, and high polarization stability. Despite advancements, power and energy scaling in short-pulsed fiber lasers continue to be constrained by two major challenges: thermally induced transverse mode instabilities (TMI) \cite{Eidam:11, Jauregui:20, zervas2019transverse, Moller:23} and Stimulated Raman Scattering (SRS) \cite{4773314}, which lead mainly to beam quality and polarization degradation \cite{zeng2023optimization, Hejaz:17}.

 Increasing the core size and reducing the fiber length are effective strategies to mitigate SRS\cite{6808413, Stutzki:14}. However, the core enlargement can lower the threshold for thermally induced TMI, which degrades beam quality and stability. Therefore, finding a balance between SRS suppression and TMI control is crucial for optimizing the performance of high-power fiber lasers\cite{Hejaz:17, Wang:23}. To tackle this balance and enhance power scaling, simultaneously reducing the complexity of laser systems, in 2008 Filippov et al. introduced the concept of tapered double-clad fiber (T-DCF). This approach opened new possibilities for direct amplification of laser pulses to high power and energy levels in a single-stage while maintaining excellent beam quality \cite{Filippov:08}. T-DCFs feature varying core and clad diameters along their length, achieved by tapering through continuous adjustments in the drawing speed. As the core size gradually increases along the fiber, the mode content remains consistent during amplification, ensuring exceptional beam quality and inherently raising the thresholds for nonlinear effects (NLEs).
 
Besides nonlinear effects, polarisation maintenance along the whole power range is another crucial aspect for high-power/energy laser systems. For T-DCFs, two primary approaches to maintaining the polarization state exist. The first involves introducing systematic strong birefringence into the fiber, which can be accomplished by incorporating stress rods, such as PANDA stress rods (so-called PT-DCF)\cite{Fedotov:18}. The second approach minimizes intrinsic birefringence by spinning the fiber preform during the drawing process. This technique disperses the preform's non-uniformities in all directions, neglecting accumulated polarization errors as the signal propagates. This method, known as spun tapered double-clad fiber (sT-DCF), has been widely adopted to enhance polarization stability at high power levels \cite{Fedotov:18, Fedotov:21, Fedotov:21-}. Both PT-DCF and sT-DCF demonstrated significant potential for efficient amplification of low-power (mW) short-pulsed signals to high power and energy levels in a single amplifier stage with excellent beam parameters \cite{Kerttula:10, Fedotov:21, Fedotov:21-, Patokoski:19, petrov2020picosecond, Fathi:24}. 
The laser with a high peak power of 1.26 MW (200 W average power) was demonstrated using PT-DCF  by A. Petrov in 2020 \cite{petrov2020picosecond}, while for high average power, a level exceeding 573 W with $\sim$ 90\% degree of linear polarisation (DOP), and slope efficiency of 75\% was achieved in sT-DCF without requiring additional pulse stretching techniques \cite{Fathi:24}. Notably, all implementations of T-DCF as the main amplifier were realized in a free-space pumping scheme, which required precise alignment to achieve high efficiency and control of heat dissipation, adding significant complexity to the setup and increasing the risk of damage. Recently, the design of completely monolithic T-DCFs using an all-fiber side-coupled combiner was reported \cite{side-pumping-mdpi-photonics-1, side-pumping-mdpi-photonics-2}. This configuration involves the non-fusion fixing of a pump-feeding fiber on a wide side surface of the T-DCF, achieving a side-pumping record of 100 W\cite{side-pumping-mdpi-photonics-2}.

In this paper, we report the amplification of various pulsed signals in a monolithic all-glass spun tapered double-clad fiber amplifier. The bulk elements-free configuration eliminates the alignment risk and results in improved performance, enabling the achievement of high average and peak power levels in a single all-fiber amplification stage by pumping near kW power.  We explore three picosecond laser systems with different repetition rates (RR) based on the monolithic all-fiber amplifier: a 1 MHz system producing 50 ps pulses with over 2 MW peak power, a 20 MHz system achieving 50 ps pulses with up to 625 W average power, and a 1 GHz system generating 20 ps pulses with up to 645 W average power. These systems exhibit excellent performance in terms of beam quality, efficiency, and polarization stability across the entire power range. Significant progress is made in amplifying short pulses directly from a few milliwatts up to 645 W average power and more than 2 MW peak power. The system maintains a high degree of linear polarisation at maximum power levels, achieving  70\%, 88.3\%, and 87.6\% for the 1MHz, 20 MHz, and 1 GHz lasers, respectively. High peak power pulses exhibit beam quality of \(M^2 \sim 2.0 \) at the highest power level with a slope efficiency of exceeding 59\%, while, high average power pulses exhibit near diffraction limited beam quality, \(M^2 <1.36 \) at the highest power level with a slope efficiency of more than 76\%. The experimental results show a remarkable step toward realizing ultra-compact, monolithic all-glass, and highly efficient high-power laser amplifier systems based on ultra-large mode area fibers, suitable for further power/energy scaling of short pulses in a wide range of repetition rates.

\section*{Experimental setup and results}
\subsection*{Experimental setup}
\begin{figure}[!t]
\centering
\includegraphics[width=\linewidth]{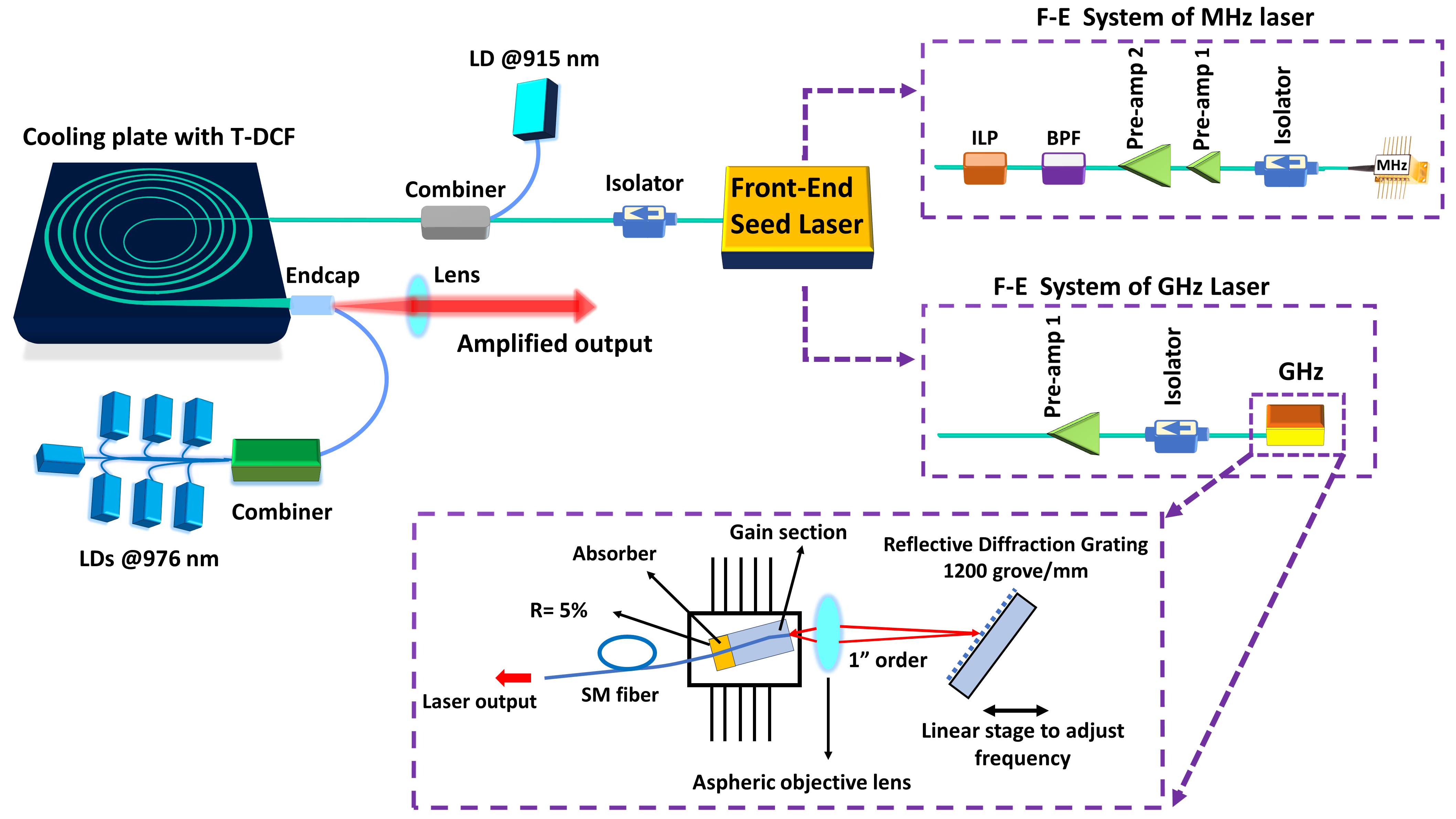}
\caption{Schematic of the experimental master oscillator-power amplifier (MOPA) architecture using monolithic all-glass sT-DCF. The components are labeled as follows: Amp, amplifier; ISO, Isolator; BPF, bandpass filter; ILP, in-line polarizer; LD, laser diode.}
\label{setup}
\end{figure}

 All amplifications of three different RR lasers were conducted using sT-DCF as the main amplifier within a master oscillator-power amplifier (MOPA) configuration. Fig. \ref{setup} illustrates a schematic of a laser system consisting of three key parts: a front-end (F-E) seed laser system, sT-DCF as a main amplifier, and a pump unit. Despite the previous studies on sT-DCF where the pump was injected in free space using multiplexer assembly\cite{Fedotov:18, petrov2020picosecond, Fedotov:21-, Fedotov:21, Fathi:24}, this work employs monolithic all-glass sT-DCF-based amplifier, enabling alignment-free kW-pumping configuration. 
 
\begin{figure}[ht]
  \begin{subfigure}{.33\textwidth}
  \centering
    \includegraphics[width=0.9\linewidth]{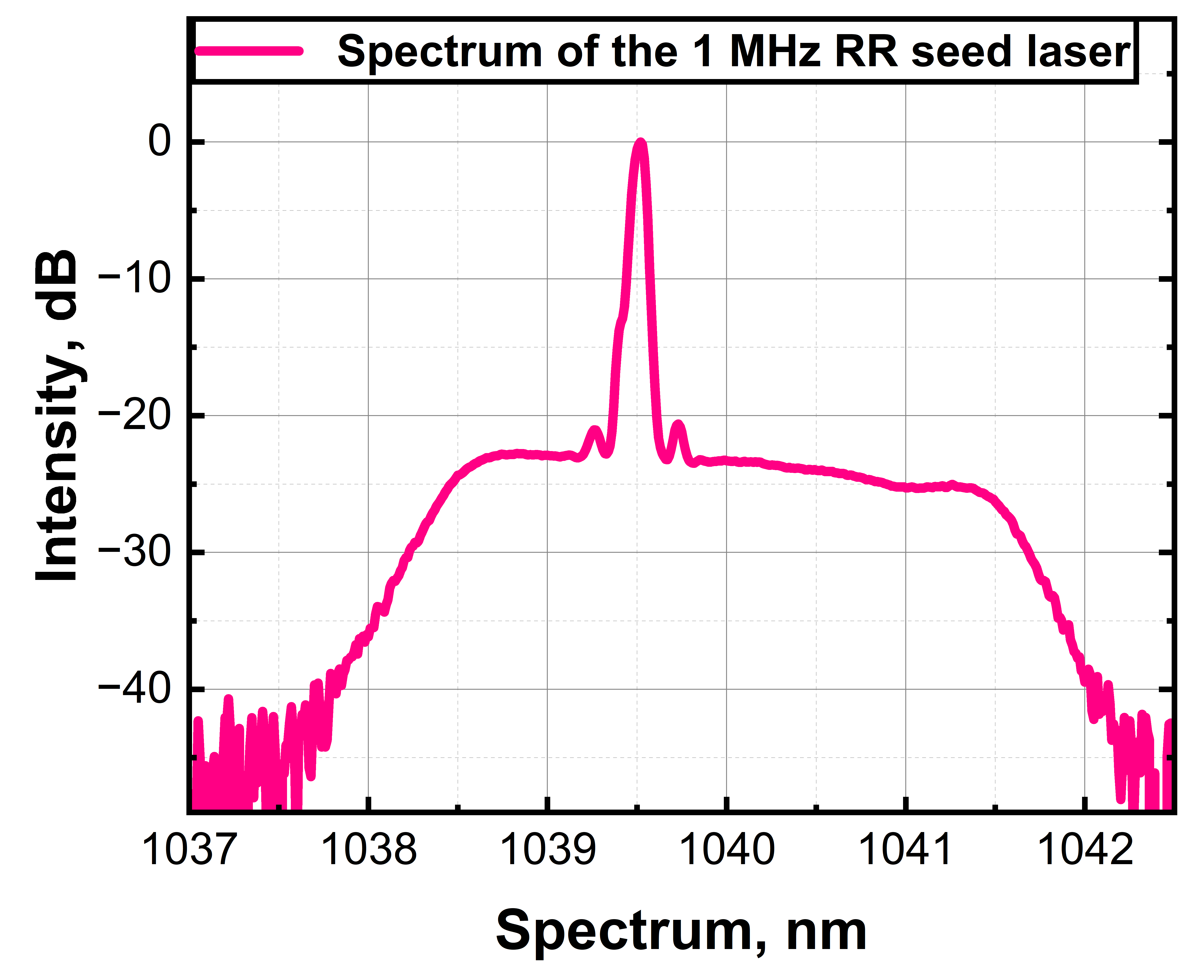}
    \caption{}
    \label{Spectrum-seed-1MHz}
  \end{subfigure}%
  \begin{subfigure}{.33\textwidth}
  \centering
    \includegraphics[width=0.9\linewidth]{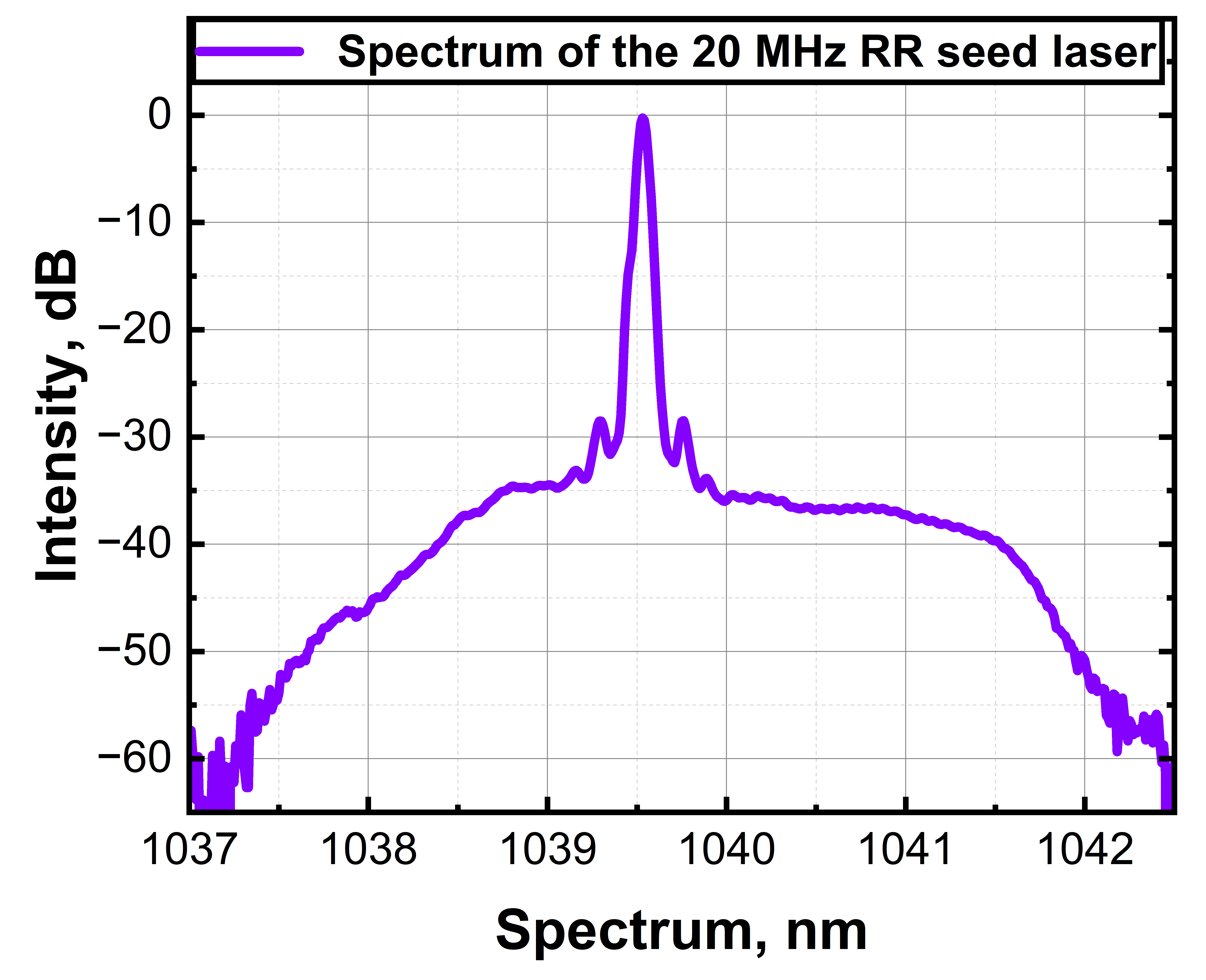}
    \caption{}
    \label{Spectrum-seed-20MHz}
  \end{subfigure}%
  \begin{subfigure}{.33\textwidth}
  \centering
    \includegraphics[width=0.9\linewidth]{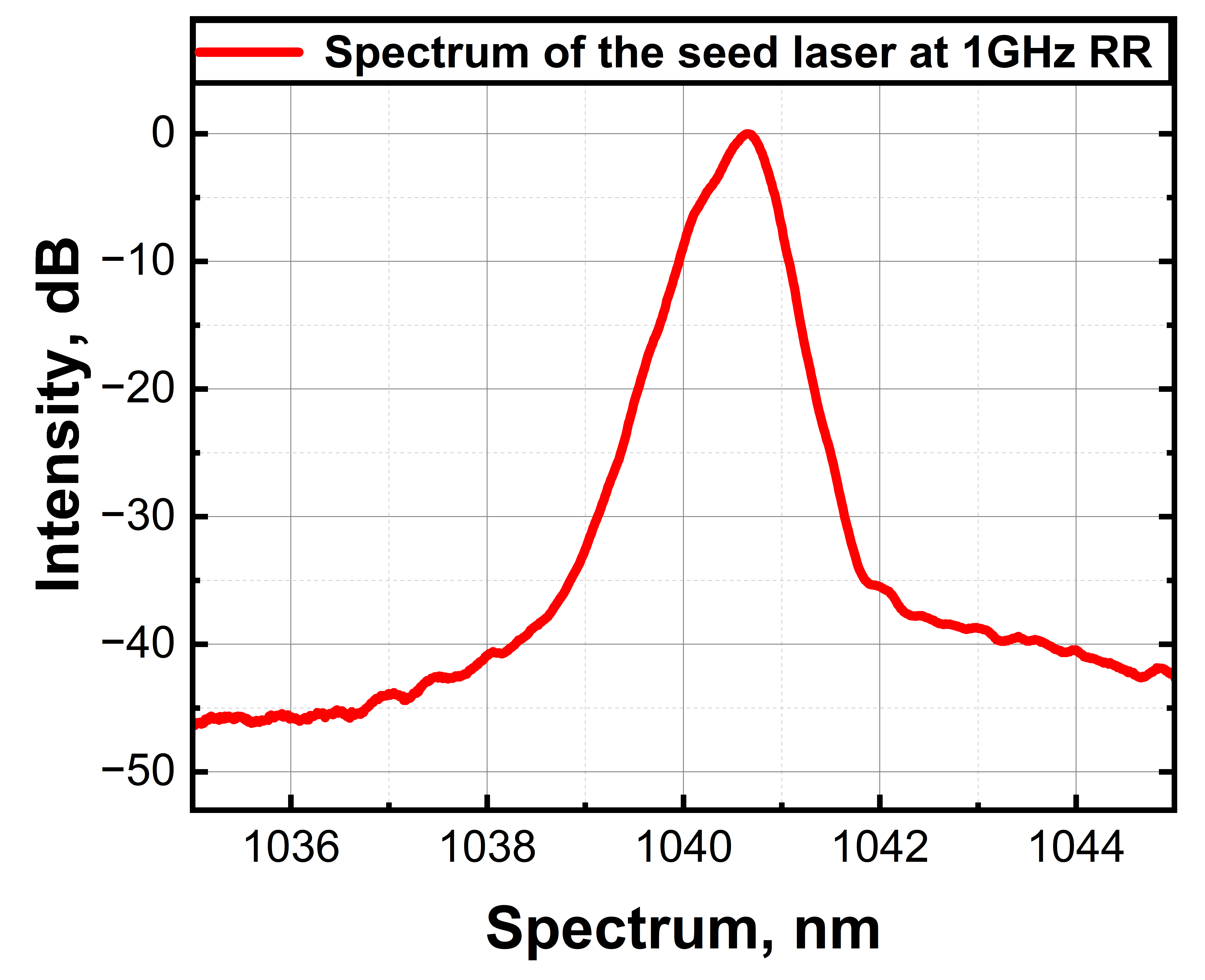}
    \caption{}
    \label{Spectrum-seed-1GHz}
  \end{subfigure}
   \caption{Optical spectrum characterization of the MHz and GHz front-end laser system. (a) at 1 MHz repetition rate with 5 mW average output power, (b) at 20 MHz repetition rate with 100 mW average output power. (c) The optical spectrum of the 20 ps pulses of 1 GHz mode-locked semiconductor seed laser with 120 mW average output power.}
   \label{seed}
\end{figure}

The F-E seed laser system employs two seed laser sources: a tunable repetition rate gain-switched (GS) laser for delivering 50 ps pulses at repetition rates (RR) of 1 MHz and 20 MHz, and a mode-locked semiconductor (ML SC) laser for generating 20 ps pulses at 1 GHz. The MHz F-E system consists of a 1040 nm GS laser diode with a polarization-maintaining (PM) fiber pigtail aligned along the slow axis, followed by two pre-amplifier stages. The GS laser generates 50 ps pulses at 1 MHz and 20 MHz, which are amplified to over 5 mW and 100 mW, respectively, using two stages of PM core-pumped single-mode Yb-doped pre-amplifiers. To ensure stable operation, the F-E system incorporates a fiber in-line optical band-pass filter (BPF) with a bandwidth of \(\pm \)2 nm to suppress amplified spontaneous emission (ASE) from the pre-amplifiers. Additionally, an in-line fiber polarizer (ILP) is included to eliminate unwanted polarization states. Fig. \ref{seed} depicts the optical spectra of MHz and GHz front-end laser systems. At 1 MHz RR, the signal-to-ASE ratio exceeds 20 dB (Fig. \ref{Spectrum-seed-1MHz}), while at 20 MHz RR, it improves to more than 28 dB (Fig. \ref{Spectrum-seed-20MHz}).

A mode-locked external cavity semiconductor (ML SC) laser is used to efficiently generate 20 ps pulses at a 1 GHz repetition rate, meeting the rising demand for high-repetition-rate lasers, particularly in burst-mode laser systems. The high repetition rate reduces the pulse's peak power, simplifying the system by eliminating the need for pulse stretchers and enabling the implementation of more compact output compressors and solid-core delivery fibers. The ML SC laser consists of a two-section chip (gain and absorber) integrated with an external cavity for efficient pulse generation. The laser is designed for high quantum efficiency (> 80\%) and low series resistance with low internal optical losses. A layout of the layers for the laser is depicted in the GHz seed box in Fig. \ref{setup}. The (In, Al) GaAs active region of the laser was grown using molecular beam epitaxy (MBE) by Innolume GmbH on the GaAs substrate. The active region of the laser consists of a single In\(_x\)Ga\(_1\)\(_-\)\(_x\)As quantum well (QW) layer to provide gain and desired wavelength. The QW layer is sandwiched between two layers of (In, Al) GaAs with the same composition to form an optical cavity in the vertical direction. This optical cavity is further sandwiched between AlGaAs cladding layers. These cladding layers support the growth of optical cavity layers by reducing stress due to lattice mismatch. 
Finally, a GaAs contact layer with p-type doping is grown on top to provide polarity to the laser diode and a minimal resistance platform for depositing the metal contacts. The external cavity length is calculated to correspond to $\sim$ 1 GHz pulse repetition rate in the ML regime.  The semiconductor laser waveguide is split into two sections to form a saturable absorber section and a gain section separated by a distance of a few micrometers. The total length of the 2-section semiconductor laser is 4 mm. A curved waveguide design is implemented to reduce back reflections in 
the external cavity configuration. The back facet on the curved end of the waveguide is coated with antireflection (AR) coatings that reflect < 0.001\% of radiation. On the other hand, the straight section (front and the output) facet on the absorber side is coated with AR coatings to reflect $\sim$ 5\% of radiation at laser wavelength around 1040 nm back into the cavity.  The output of the laser is butt-coupled to the PM single-mode fiber. The external cavity of the laser is set up in the Littrow configuration. The output from the curved side of the waveguide was focused on a 1200 grooves/mm blaze reflective diffraction grating using an aspherical lens (f = 4.5 mm).  The reflectivity of the grating at 1µm wavelength in the 1st order is ~80\%. The grating is aligned to reflect the 1st order into the gain section of the laser. Aligning the diffraction grating enables both wavelength tuning and stabilization of the laser, as well as adjustment of the repetition rate in the mode-locked regime. By carefully adjusting the grating, the laser is set to operate at 1040 nm with a 1 GHz pulse repetition rate. Fig. \ref{Spectrum-seed-1GHz} depicts the optical spectrum of the 20 ps pulses of a 1 GHz mode-locked semiconductor seed laser (more than 40 dB of signal-to-noise ratio).

At the end of both the MHz and GHz F-E laser systems, a 10 W PM isolator protects the seed laser from high-power back reflections. This isolator has a minimum extinction ratio of 20 dB and a minimum isolation of 40 dB across all polarization states. The output of the F-E system was then spliced to the input signal port of a (2+1)\(\times \) 1 combiner, which is pumped by an 18 W laser diode operating at 915 nm. The output port of the combiner is then spliced to the narrower side of the sT-DCF amplifier.

\begin{figure}[!t]
\centering
\includegraphics[width=0.45\linewidth]{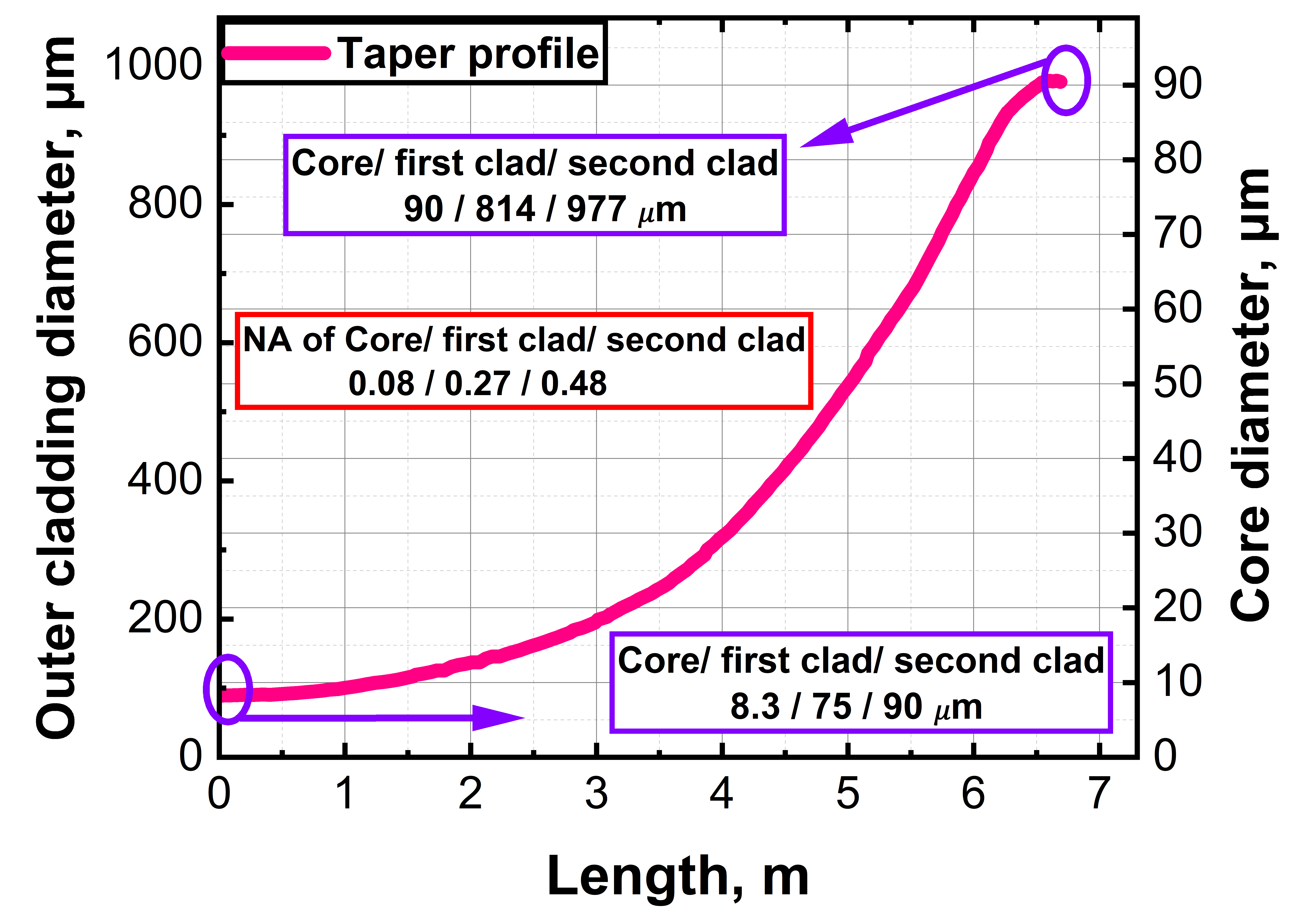}
\caption{Longitudinal profile of the active spun tapered-double clad fiber.}
\label{Taper}
\end{figure}

A spun Yb-doped tapered double-clad fiber is utilized as the main gain medium for the amplification of both MHz and GHz repetition rate laser systems. This fiber features two key attributes; tapering and spinning. Tapering is achieved through continuous adjustments in the drawing speed of the fiber preform, creating a conical shape of the fiber. However, the spinning is accomplished by rotating the preform at a specific angular velocity during the drawing process, which uniformly disperses preform non-uniformities in all directions and mitigates polarization errors as the signal propagates \cite{Fedotov:21-}. Consequently, the intrinsic internal birefringence is reduced to as low as $\approx$ \(10^- \)\(^8 \), making the state of polarization (SOP) insensitive to heating caused by quantum defects \cite{Fedotov:21, Fedotov:21-}. Spinning the fiber affects the mode content and pump absorption. Therefore, optimizing the pitch length, core parameters, and cladding shape is crucial\cite{Fedotov:21-}. The sT-DCF utilized in this work has a constant spinning pitch length of 50 mm. The modified chemical vapor deposition (MCVD) technique was employed to manufacture the active core rod. The necessary thickness of the first cladding (pure silica) was created by sleeving with F300 silica tubes on a glass-working lathe. The fluorine-doped second cladding was deposited with the help of low-pressure microwave plasma\cite{Grishchenko_2017}. Numerical Apertures (NA) for core/first cladding and first cladding/second cladding are 0,08 and 0,27 respectively. The cross-section of the first cladding was shaped to double D geometry to enhance absorption of pump irradiation. Glass of the sT-DCF was protected by a pair of acrylate coatings, where a low-index primary polymer creates an extra reflective layer with NA=0,48 relative to the first cladding. The taper exhibits an 800 dB/m core absorption at 976 nm. The diameters of the core, first clad, and second clad of the sT-DCF range from 8.3/75/90 \(\mu \)m to 90/814/977 \(\mu \)m over a length of $\approx$ 6.7 m length. Additionally, the sT-DCF features a double D-shape geometry on its first clad (pure silica glass) to enhance pump absorption and is protected by a low-index polymer coating.  The longitudinal profile of the active spun taper is illustrated in Fig. \ref{Taper}.
The sT-DCF is coiled with a 35 cm diameter and positioned in the grooves of a water-cooled aluminum plate to ensure efficient heat dissipation. A 2-degree angled silica glass endcap (3 mm in diameter and 6 mm in length) is spliced to the wide end of the sT-DCF to eliminate back reflections and prevent potential damage from the high-power density at the air/silica interface. The amplifier's pump fiber is spliced to the endcap facet.

The main amplifier's pump unit comprises seven wavelength-stabilized laser diodes, each delivering up to 120 W of output power at 976 nm, providing a total output pump power of over 840 W. The outputs from the diodes are combined using a 7x1 fused-fiber pump combiner, which couples the pump light into a single fiber with a 200 \(\mu \)m core diameter. The output port of the combiner is then spliced to the pump fiber of sT-DCF, injecting the pump light into the wide side of the sT-DCF in a backward pump configuration. This alignment-free and integrated pumping scheme simplifies the system, reducing its complexity and making it more compact. The sT-DCF is predominantly pumped from the wide side using a 976 nm pump source, with an additional 18 W of power at 915 nm injected from the narrow side of the taper through a (2+1)\(\times \)1 combiner. This dual-sided pumping scheme ensured optimal gain distribution and efficient energy transfer along the entire taper length, enabling the amplification of high-peak-power pulses while maintaining a compact system architecture. The approach supports effective amplification by maximizing pump absorption and minimizing nonlinear effects throughout the fiber.

\subsection*{Amplification of 1 MHz laser}
\begin{figure}[!t]
  \begin{subfigure}{.33\textwidth}
  \centering
    \includegraphics[width=0.935\linewidth]{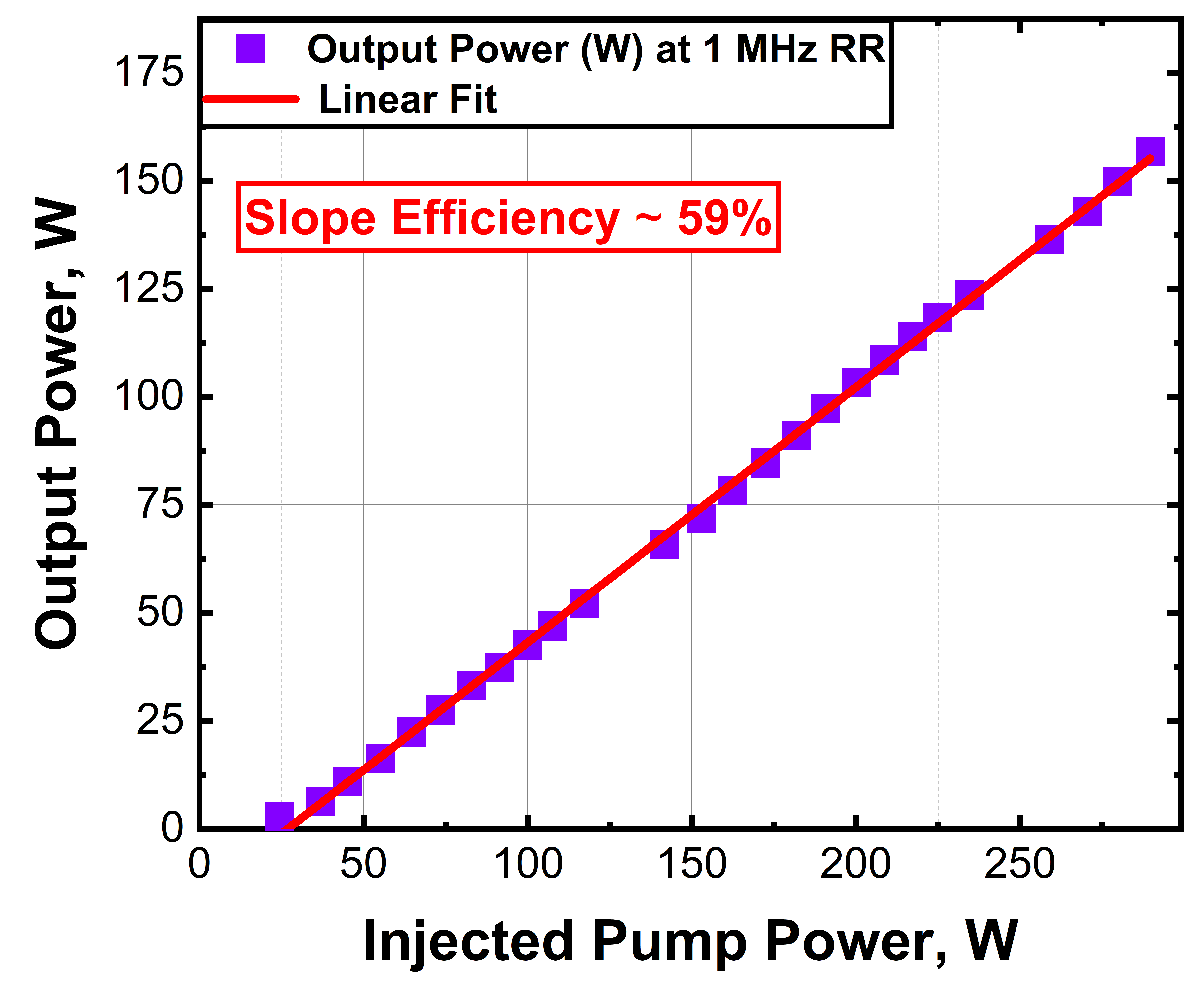}
    \caption{}
    \label{Slope-1MHz}
  \end{subfigure}%
  \begin{subfigure}{.33\textwidth}
  \centering
    \includegraphics[width=0.96\linewidth]{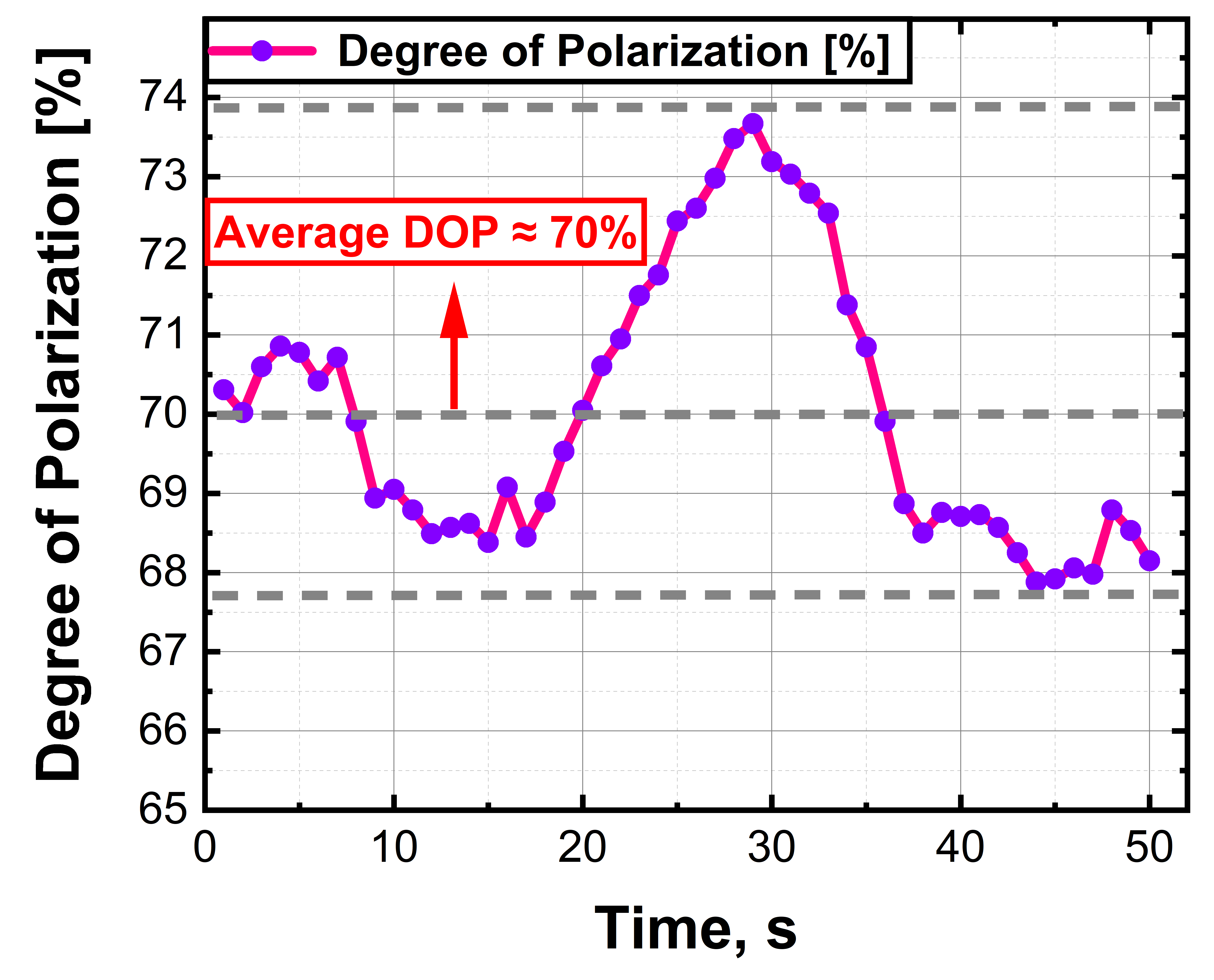}
    \caption{}
    \label{DOP-1MHz}
  \end{subfigure}%
   \begin{subfigure}{.33\textwidth}
  \centering
    \includegraphics[width=0.98\linewidth]{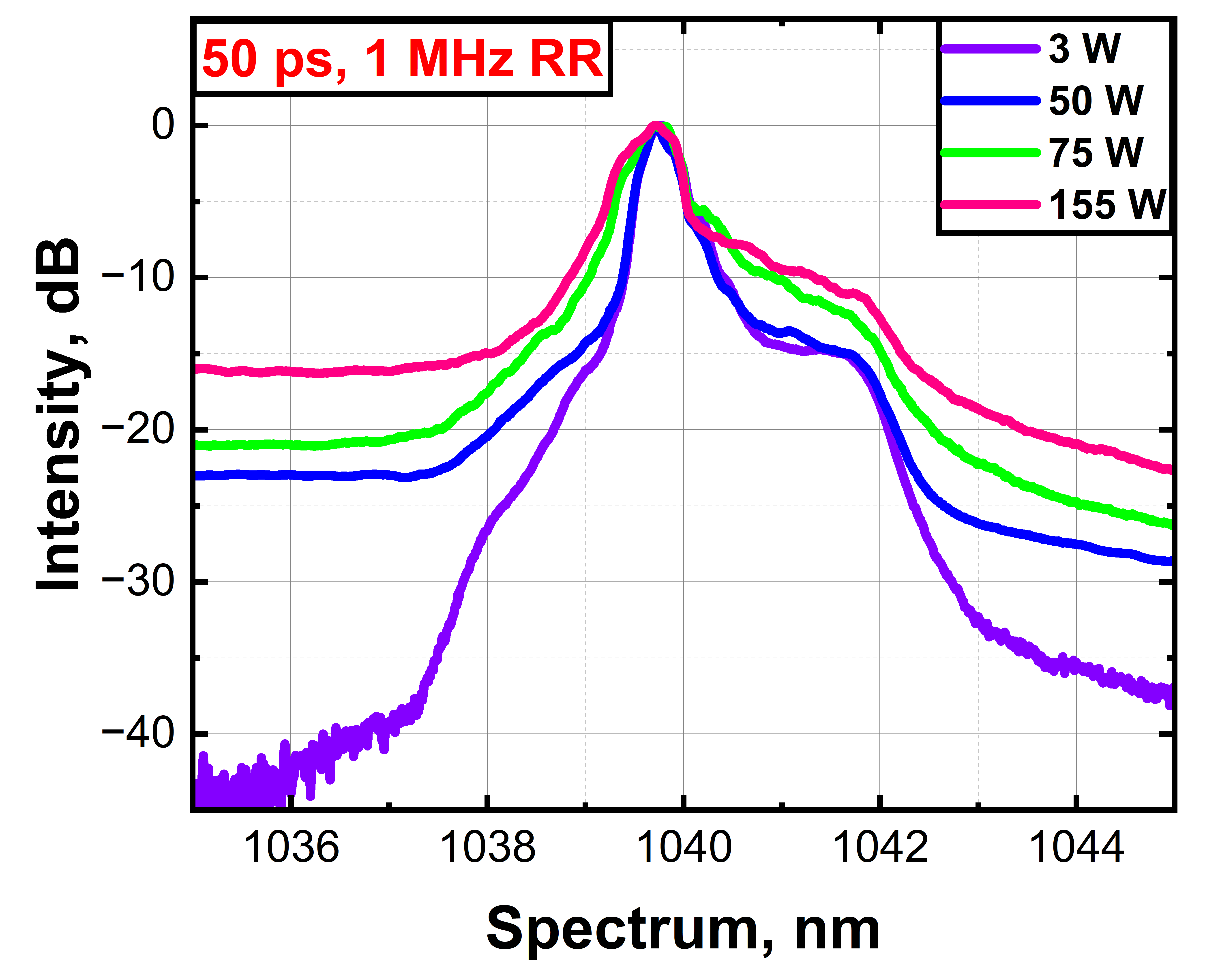}
    \caption{}
    \label{Spectrum-Narrow-1MHz}
  \end{subfigure}
  \begin{subfigure}{.33\textwidth}
  \centering
    \includegraphics[width=0.95\linewidth]{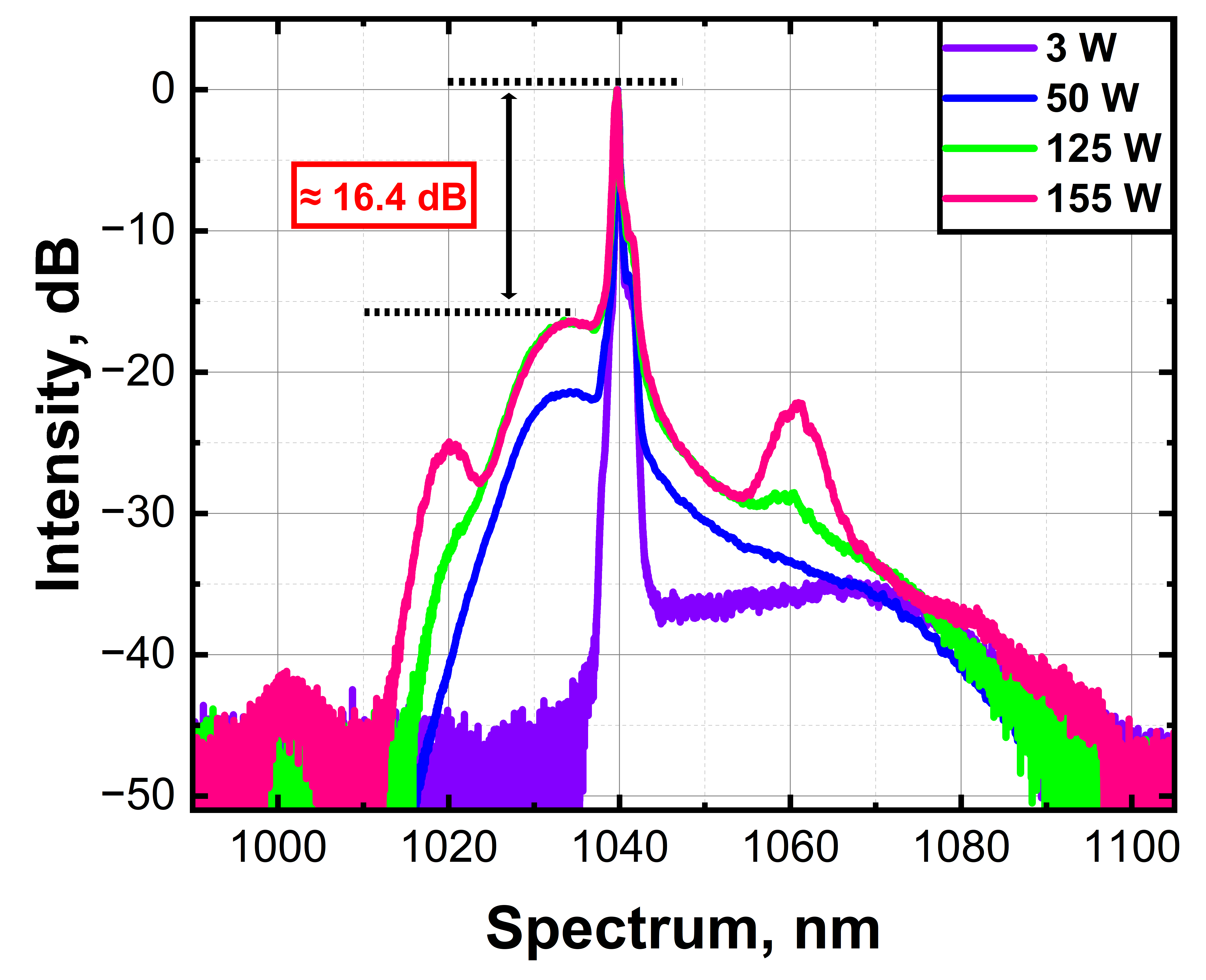}
    \caption{}
    \label{Spectrum-Wide-1MHz}
  \end{subfigure}%
  \begin{subfigure}{.33\textwidth}
  \centering
    \includegraphics[width=0.99\linewidth]{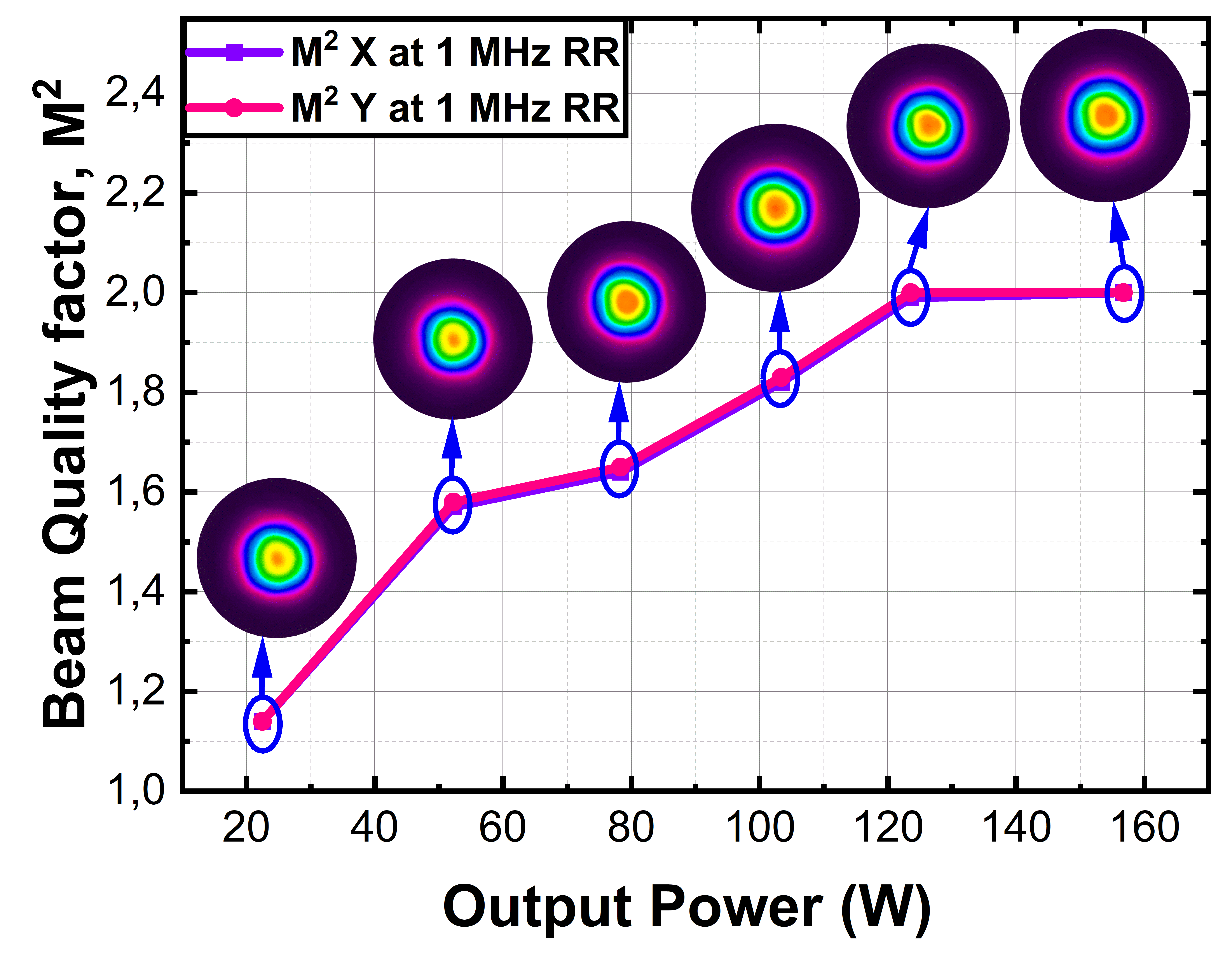}
    \caption{}
    \label{M2evolution-1MHz}
  \end{subfigure}%
  \begin{subfigure}{.33\textwidth}
  \centering
    \includegraphics[width=1\linewidth]{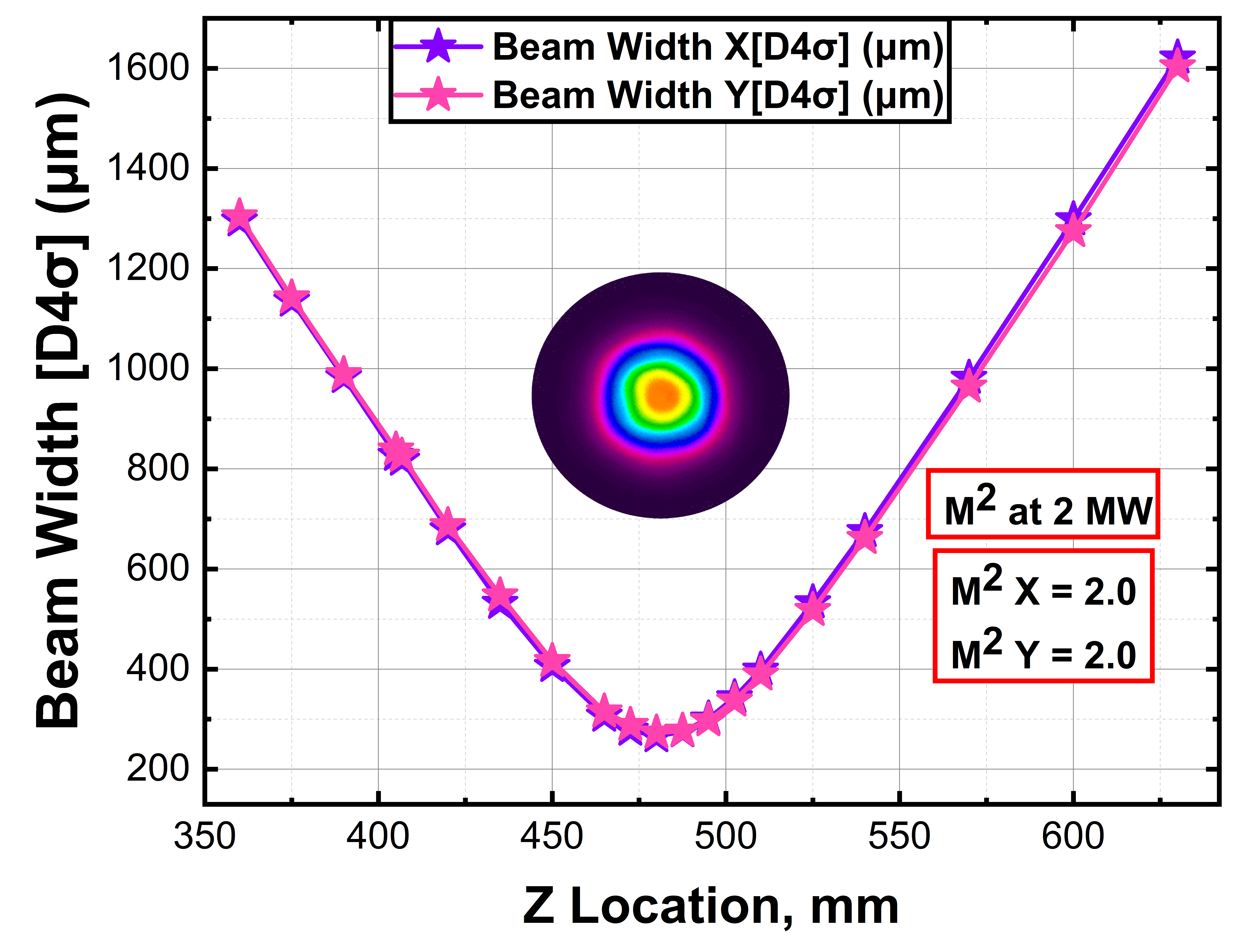}
    \caption{}
    \label{M2curve-1MHz}
  \end{subfigure}
  \caption{Experimental characterization of the amplified 1 MHz pulsed signal. (a) The output power of the 1 MHz laser system versus the total launched pump power, showing the power scaling performance. (b) Degree of polarization measurements at the maximum output power. (c) and (d) represent the output optical spectrums at different output powers in narrow and wide spectral ranges, respectively. (e) The beam quality factor (\(M^2\)) measurements of the 1 MHz laser system at different output power alongside corresponding beam profiles. (f) Beam quality factor (\(M^2\)) measurement with beam profile inset at 2 MW peak power.}
   \label{1 MHz}
\end{figure}

This section presents the amplification of 50 ps pulses at a 1 MHz repetition rate. The direct amplification of low repetition rates picosecond pulses is limited in standard constant core/clad ration fibers due to the onset of strong non-linear effects such as self-phase modulation and four-wave mixing, particularly at high power. However, T-DCFs offer the solution through their unique geometrical configurations for minimizing nonlinear interactions and achieving high power amplification.
Utilizing all-glass sT-DCF, over 2 MW peak power (155 W average power) is achieved by direct amplification through sT-DCF, with ~59\% optical efficiency. The peak power was determined by considering only the signal region, excluding the amplified spontaneous emission contribution from the total output power. The dual-sided pumping strategy contributed to achieving the high peak powers while ensuring good energy transfer and gain distribution along the fiber. However, the fast growth of ASE and degradation of DOP limit further power scaling. Fig. \ref{Slope-1MHz} depicts the average power at the laser system output as a function of the injected pump power. At maximum average output power (2 MW peak power), the system demonstrates around 70\% of the degree of polarization (DOP), as shown in Fig. \ref{DOP-1MHz}. The polarization was measured over a 50-second interval using a commercial polarimeter (PAX1000IR2/M). This assessment ensures the system maintains a reasonable polarization quality even at high peak power levels.

Fig. \ref{Spectrum-Narrow-1MHz} illustrates the optical spectra of the amplified signals across a narrow spectral range for different power levels. It shows that the 3 dB spectral bandwidth of the signal broadens slightly as the output power increases, from 0.45 nm at 3 W to 0.67 nm at 155 W.
The optical spectra of the amplified signals across the wide wavelength range for different power levels are presented in Fig. \ref{Spectrum-Wide-1MHz}. The spectra reveal a 16.4 dB amplified spontaneous emission (ASE) background but no evidence of stimulated Raman scattering (SRS), which typically poses a challenge in high-power fiber systems. However, the appearance of new spectral sidebands around 1020 nm and 1060 nm (signal at 1040 \(\pm \) 20 nm), likely due to Polarization modulation instability (PMI)\cite{Murdoch:95}, highlights the impact of nonlinearities at elevated peak powers. These PMI-induced sidebands are generated by the coherent interaction of two polarization modes in a weakly birefringent optical fiber, potentially distorting the optical spectrum, degrading beam quality, and reducing overall system efficiency.

In terms of beam quality, \(M^2\) factor measurements provide valuable insights, see Fig. \ref{M2evolution-1MHz} and Fig. \ref{M2curve-1MHz}. The measurements were carried out according to the ISO 11146 standard using the 4$\sigma$-method. Fig. \ref{M2evolution-1MHz} shows the evolution of beam quality with increasing output power, along with corresponding beam profiles. Notably, the intensity beam profile remained highly symmetric with a single intensity maximum across a wide range of operating conditions, even at maximum output power. At maximum power, the beam quality measured as \(M^2X/M^2Y=2.0/2.0\), indicating no sign of mode instability; see Fig. \ref{M2curve-1MHz} (insets: near-field beam profile of the beam at 2 MW peak power). We attribute the increase in \(M^2\) value with the output power growth mainly to the PMI effect and new wavelength generation at orthogonal polarization states.

The experimental results of amplifying high-peak-power pulses at a 1 MHz repetition rate via all-glass sT-DCF demonstrate effective amplification while maintaining good beam quality, high polarization stability, and manageable nonlinearities. The results mark a significant step toward scaling up short-pulse amplification in compact, high-power fiber laser systems.

\subsection*{Amplification of 20 MHz laser}
\begin{figure}[!t]
  \begin{subfigure}{.33\textwidth}
  \centering
    \includegraphics[width=0.935\linewidth]{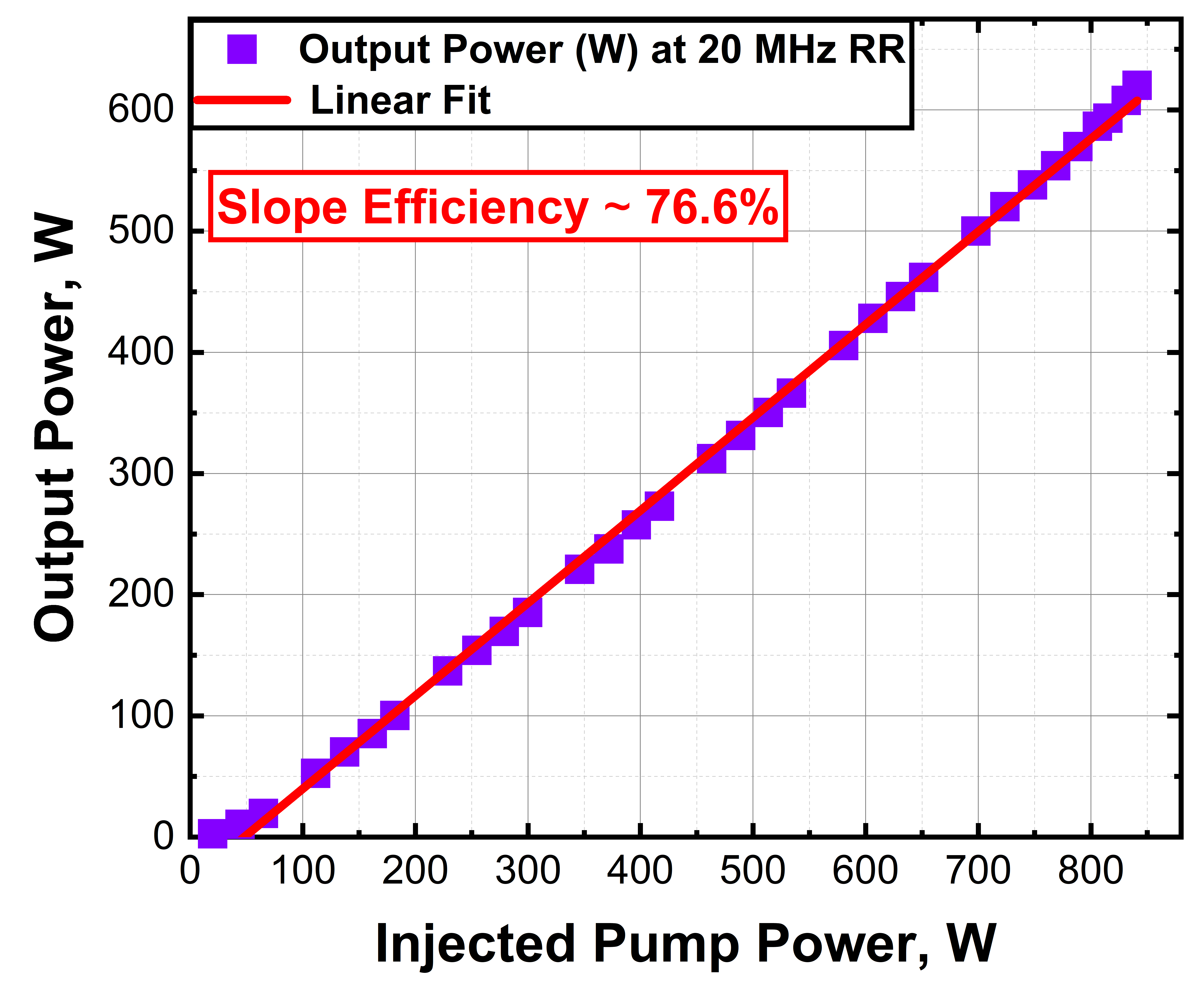}
    \caption{}
    \label{Slope-20MHz}
  \end{subfigure}%
  \begin{subfigure}{.33\textwidth}
  \centering
    \includegraphics[width=1\linewidth]{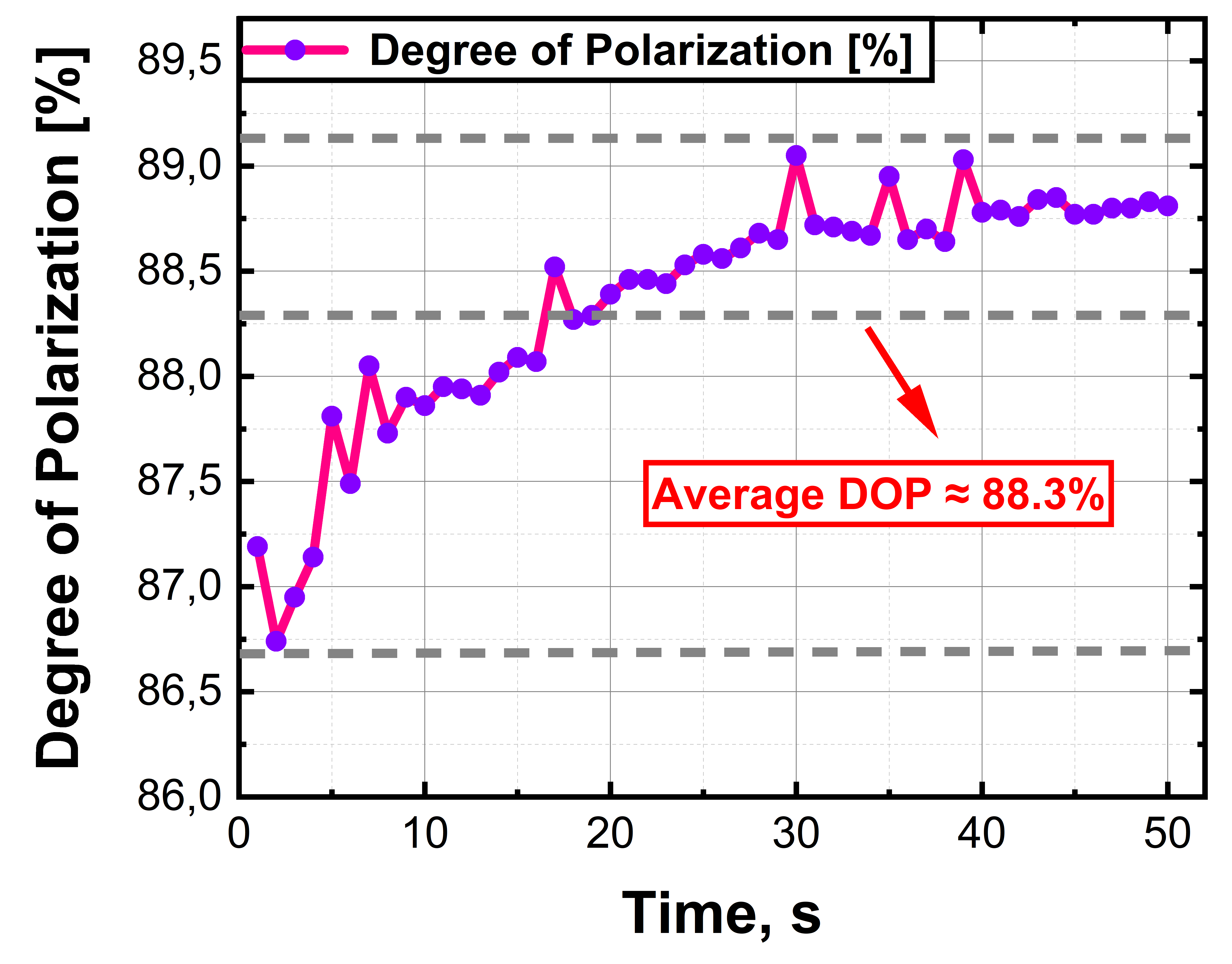}
    \caption{}
    \label{DOP-20MHz}
  \end{subfigure}%
   \begin{subfigure}{.33\textwidth}
  \centering
    \includegraphics[width=0.95\linewidth]{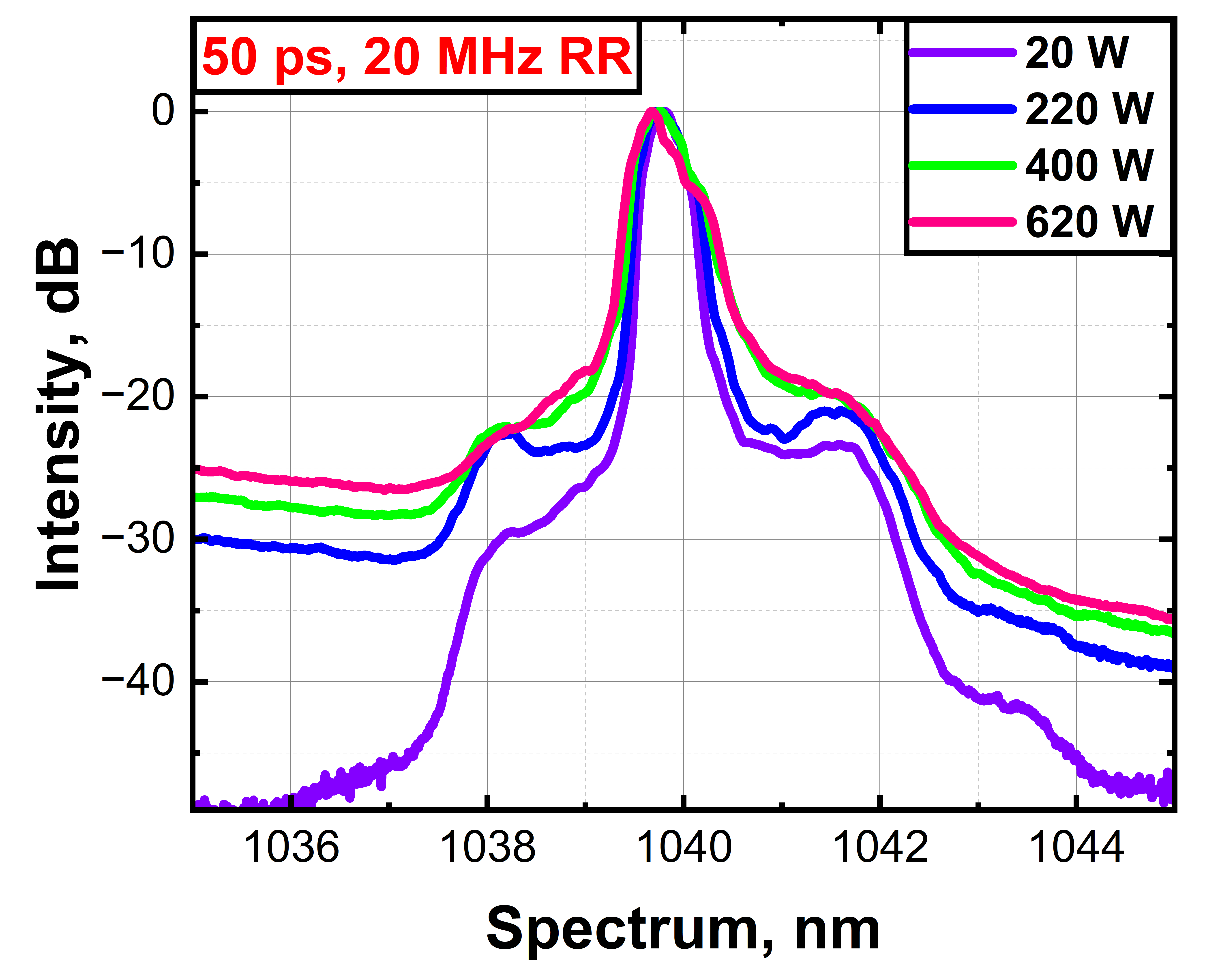}
    \caption{}
    \label{Spectrum-Narrow-20MHz}
  \end{subfigure}
  \begin{subfigure}{.33\textwidth}
  \centering
    \includegraphics[width=0.93\linewidth]{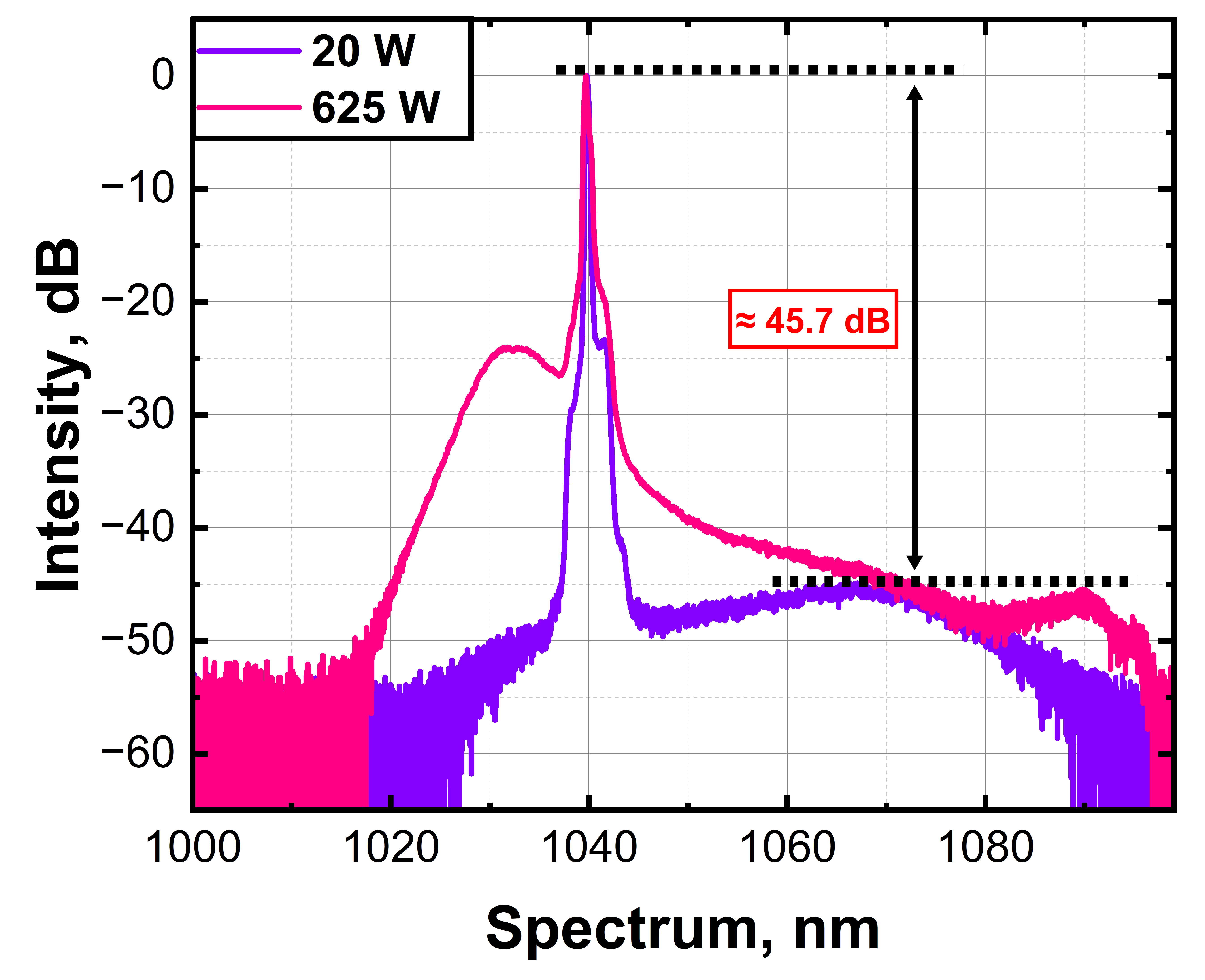}
    \caption{}
    \label{Spectrum-Wide-20MHz}
  \end{subfigure}%
  \begin{subfigure}{.33\textwidth}
  \centering
    \includegraphics[width=1\linewidth]{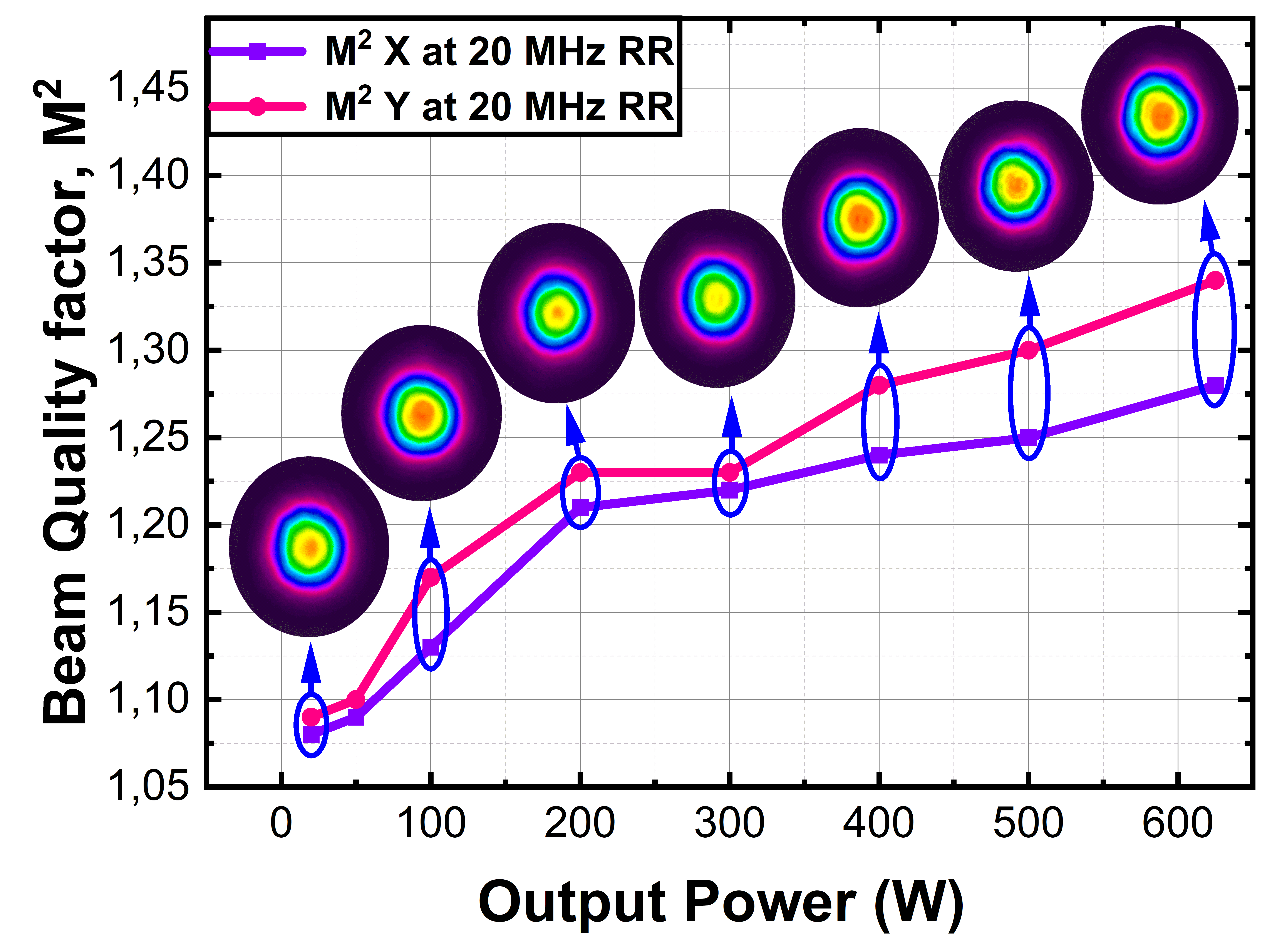}
    \caption{}
    \label{M2evolution-20MHz}
  \end{subfigure}%
  \begin{subfigure}{.33\textwidth}
  \centering
    \includegraphics[width=0.94\linewidth]{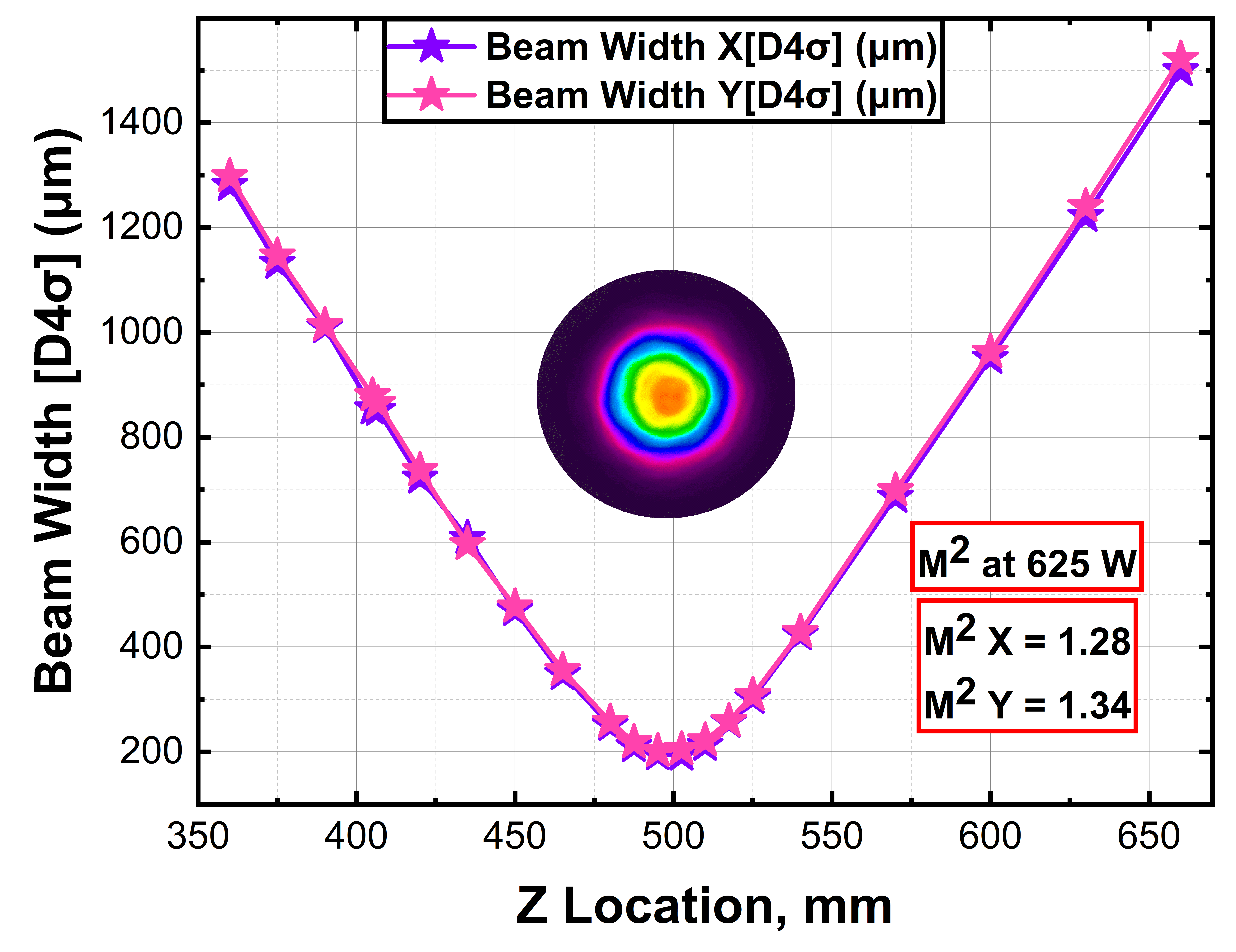}
    \caption{}
    \label{M2curve-20MHz}
  \end{subfigure}
  \caption{Experimental characterization of the amplified 20 MHz pulsed signal. (a) The output power of the 20 MHz laser system versus the total launched pump power, showing the power scaling performance. (b) Degree of polarization measurements at the maximum output power. (c) and (d) represent the output optical spectrums at different output powers in narrow and wide spectral ranges, respectively. (e) The beam quality factor (\(M^2\)) measurements of 20 MHz laser system at different output power alongside corresponding beam profiles. (f) Beam quality factor (\(M^2\)) measurement with beam profile inset at 625 W average output power.}
   \label{20MHz}
\end{figure}

This section presents the amplification of 50 ps pulses at a 20 MHz repetition rate in the all-glass sT-DCF amplifier module. Operating at full pump power, 625 W average output power is achieved with a slope efficiency of 76.6\%. Further power scaling for the amplification of pulses was limited by the available pump power. Fig. \ref{Slope-20MHz} depicts the average output power versus the injected pump power. At the maximum output power of 625 watts, the degree of polarization was measured as 88.3\%. This high degree of polarization signifies that the system successfully maintains a superior level of linear polarization even under the high thermal loads associated with maximum output power, see Fig. \ref{DOP-20MHz}.
Fig. \ref{Spectrum-Narrow-20MHz} illustrates the optical spectra of the amplified signals across a narrow spectral range for different power levels. It shows that the 3 dB spectral bandwidth of the signal broadens slightly as the output power increases, from 0.38 nm at 20 W to 0.43 nm at 625 W.
The optical spectra of the amplified signals across a wide spectrum range for different power levels are shown in Fig. \ref{Spectrum-Wide-20MHz}. The spectra demonstrate a low amplified spontaneous emission (ASE) background of $\approx$ 24.3 dB and a high signal-to-Raman peak suppression ratio of $\approx$ 45.7 dB, indicating efficient direct amplification of picosecond pulses.

Fig. \ref{M2evolution-20MHz}, and \ref{M2curve-20MHz} present the beam quality characterization of the 20 MHz laser system based on \(M^2\) factor measurement. The evolution of the beam quality with increasing the output power, along with the corresponding beam profiles, is depicted in Fig.\ref{M2evolution-20MHz}. At maximum power, the beam quality measured as \(M^2X/M^2Y=1.28/1.34\), indicating no sign of mode instability; see Fig. \ref{M2curve-20MHz} (insets: near-field beam profile of the beam at 625 W average power).

\subsection*{Amplification of 1 GHz laser}
\begin{figure}[!t]
  \begin{subfigure}{.33\textwidth}
 \centering
    \includegraphics[width=0.95\linewidth]{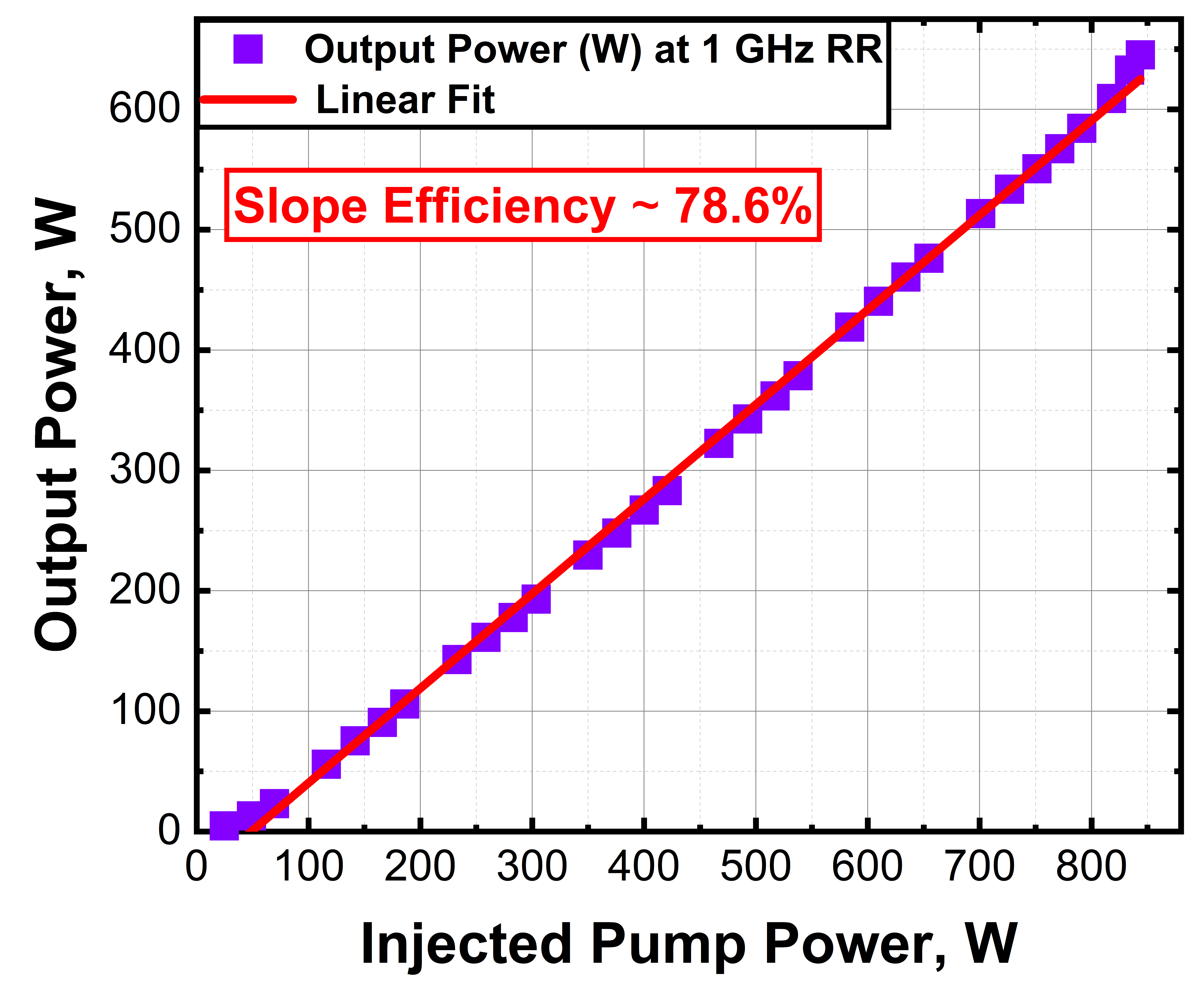}
    \caption{}
    \label{Slope-1GHz}
  \end{subfigure}%
  \begin{subfigure}{.33\textwidth}
  \centering
    \includegraphics[width=1\linewidth]{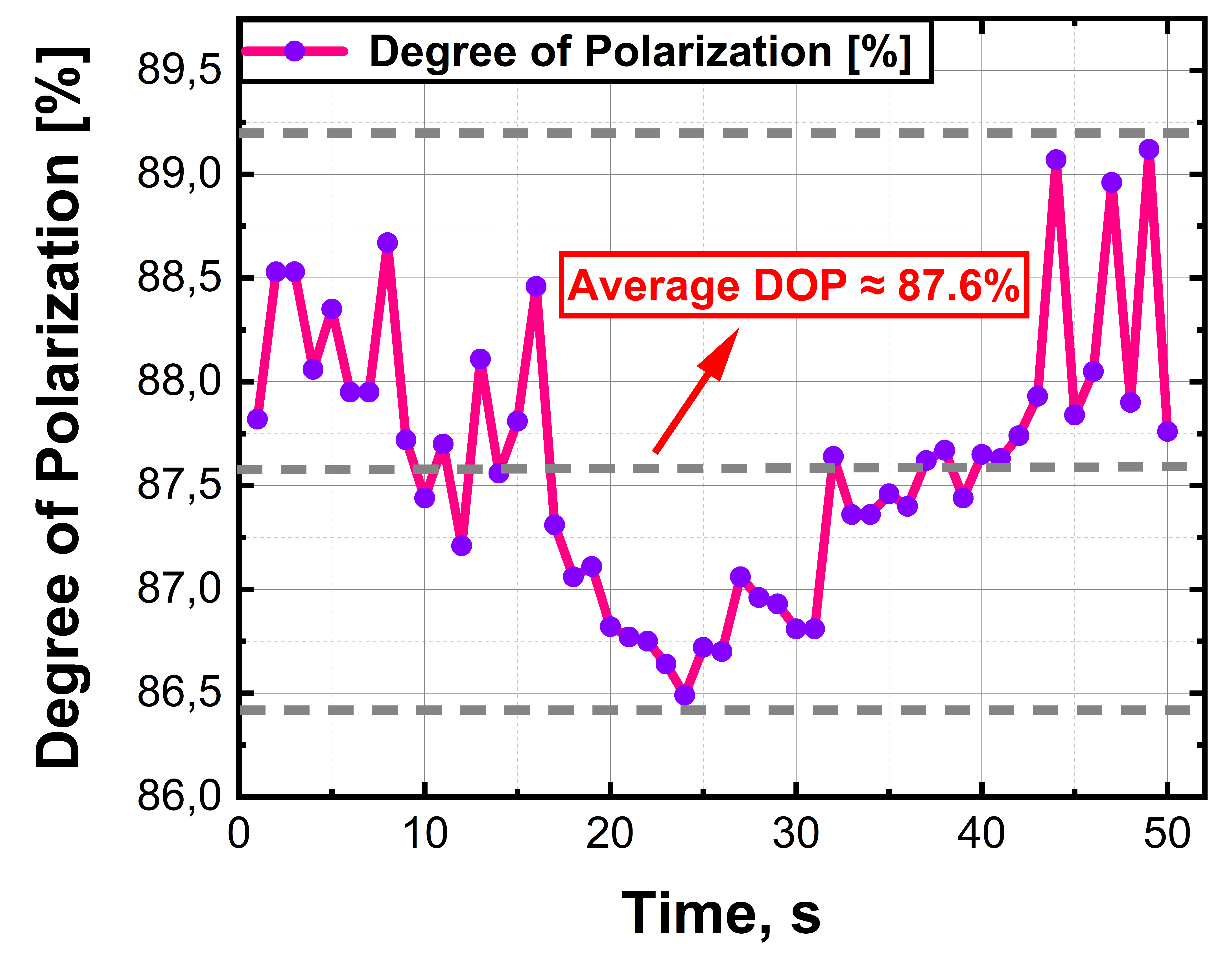}
    \caption{}
    \label{DOP-1GHz}
  \end{subfigure}%
   \begin{subfigure}{.33\textwidth}
  \centering
    \includegraphics[width=0.98\linewidth]{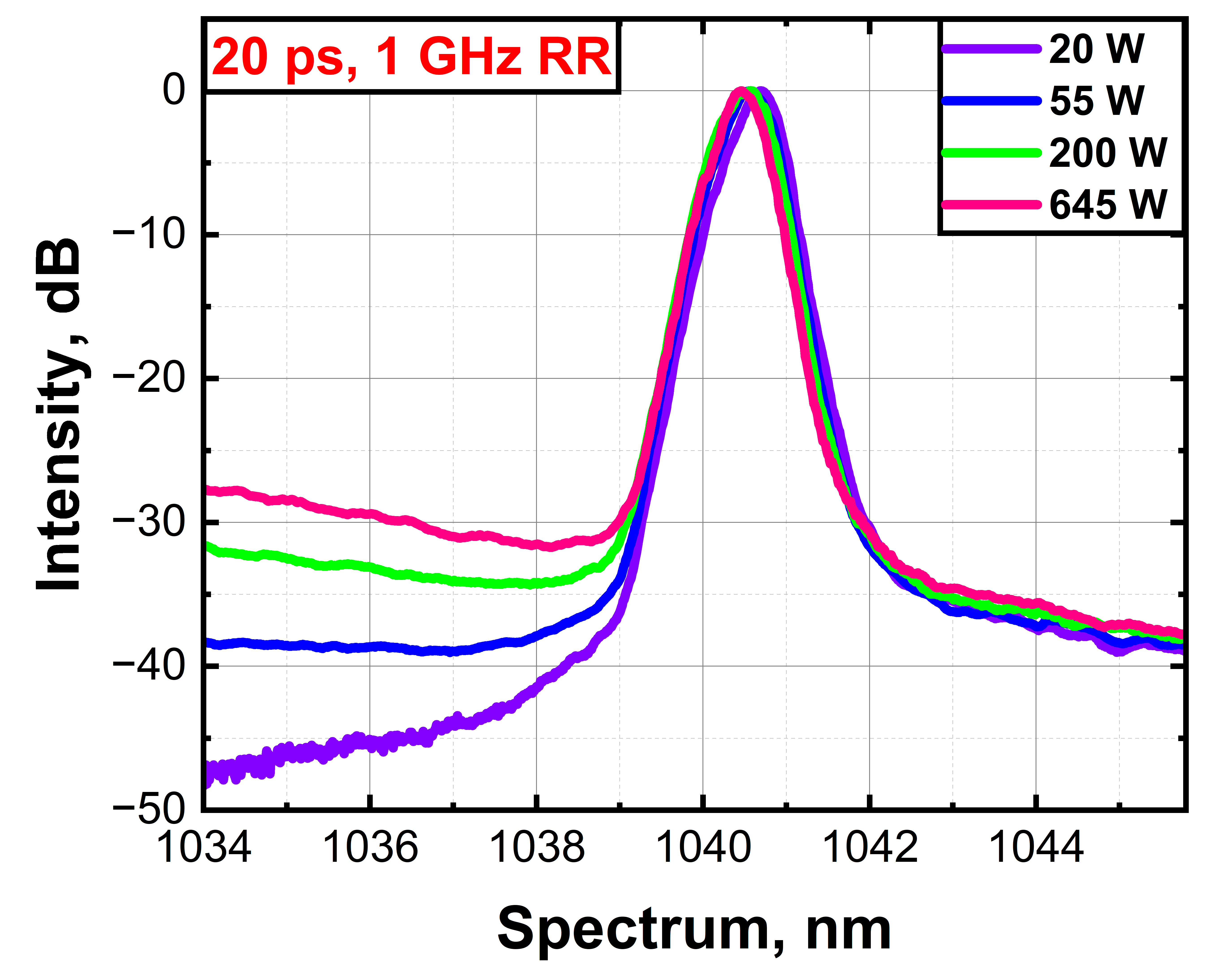}
    \caption{}
    \label{Spectrum-Narrow-1GHz}
  \end{subfigure}
  \begin{subfigure}{.33\textwidth}
  \centering
    \includegraphics[width=0.93\linewidth]{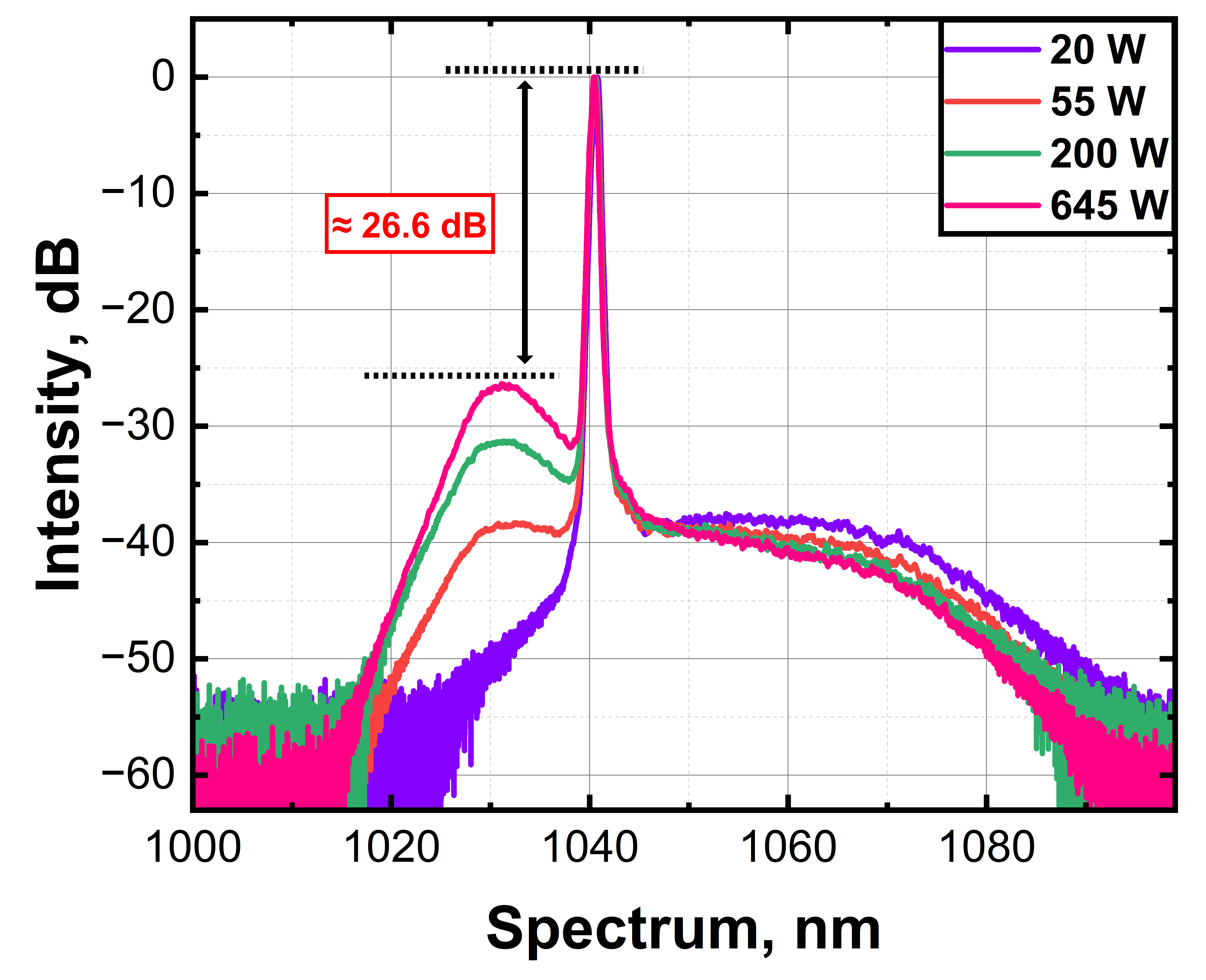}
    \caption{}
    \label{Spectrum-Wide-1GHz}
  \end{subfigure}%
  \begin{subfigure}{.33\textwidth}
  \centering
    \includegraphics[width=1\linewidth]{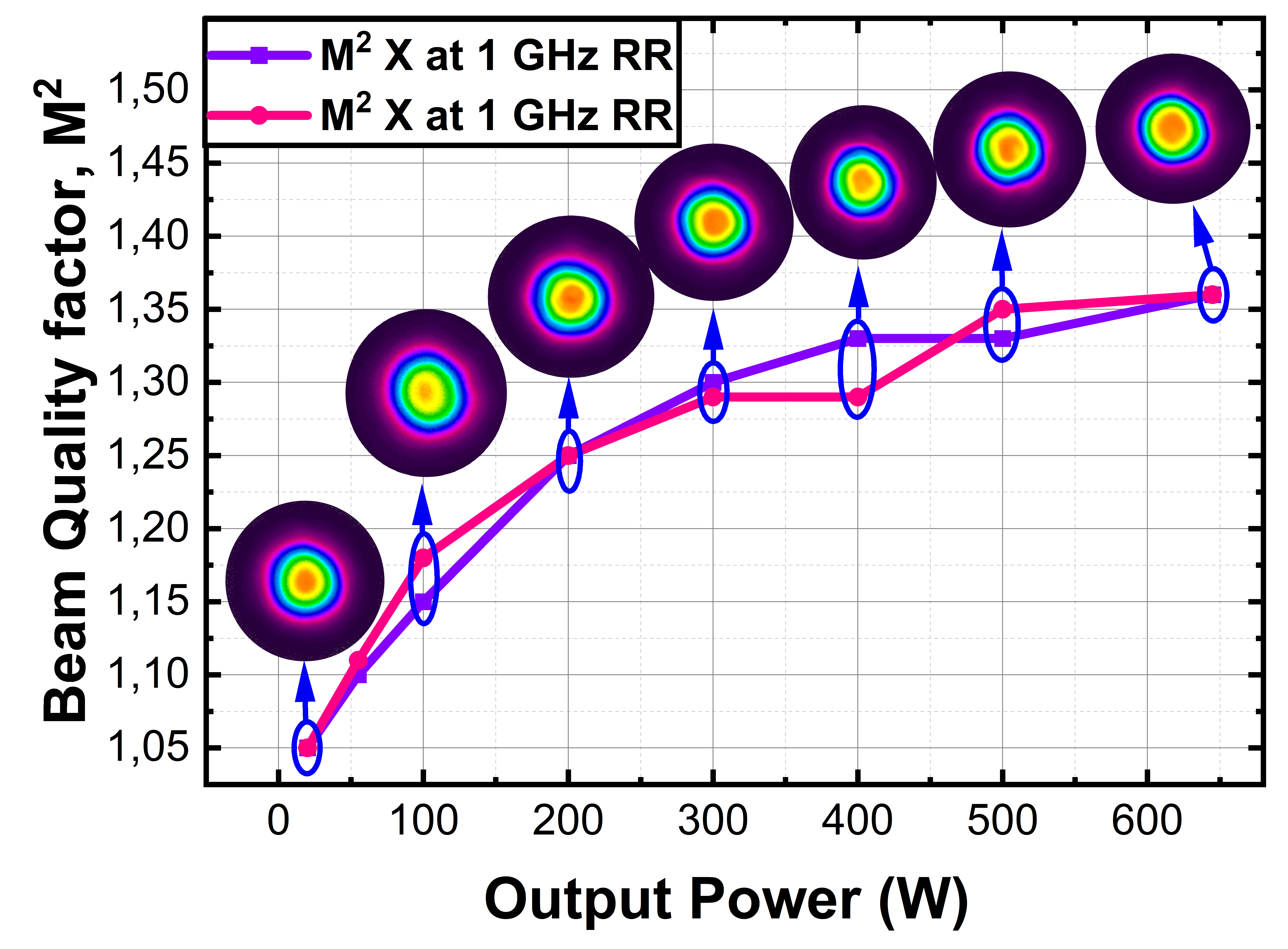}
    \caption{}
    \label{M2evolution-1GHz}
  \end{subfigure}%
  \begin{subfigure}{.33\textwidth}
  \centering
    \includegraphics[width=0.94\linewidth]{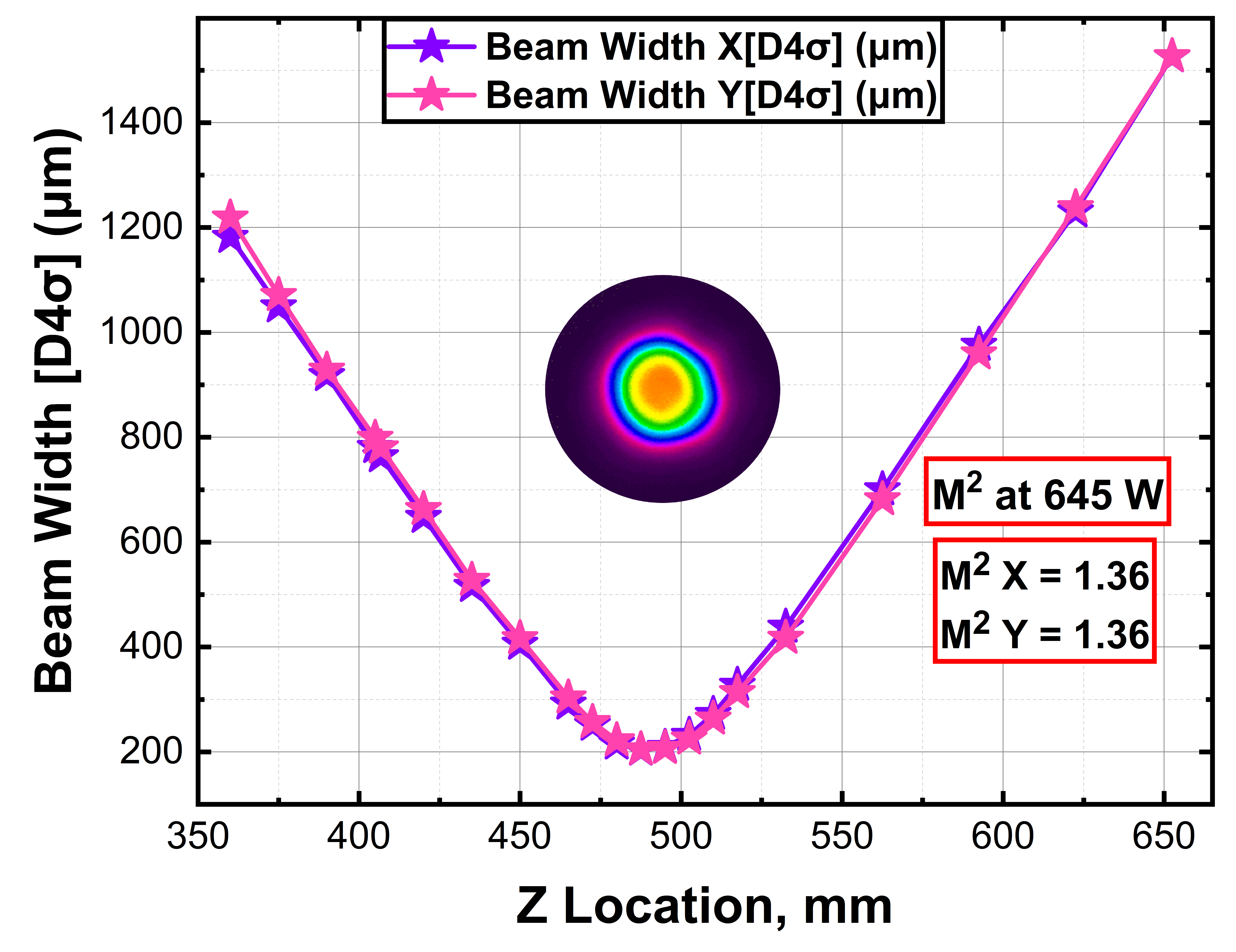}
    \caption{}
    \label{M2curve-1GHz}
  \end{subfigure}
  \caption{Experimental characterization of the amplified 1 GHz laser system. (a) The output power of the 1 GHz laser system versus the total launched pump power, showing the power scaling performance. (b) Degree of polarization measurements at the maximum output power. (c) and (d) represent the output optical spectrums at different output powers in narrow and wide spectral ranges, respectively. (e) The beam quality factor (\(M^2\)) measurements of the 1 GHz laser system at different output power alongside corresponding beam profiles. (f) Beam quality factor (\(M^2\)) measurement with beam profile inset at 645 W average output power.}
   \label{1 GHz}
\end{figure}

This section reports on the successful amplification of 20 ps pulses at a high repetition rate of 1 GHz using an all-glass sT-DCF laser amplifier. The system achieved an average output power of 645 W with a slope efficiency of 78.6\%. The available pump power ultimately limited further power scaling, as shown in Fig. \ref{Slope-1GHz}. Polarization measurements revealed a degree of polarization (DOP) of 87.6\%, highlighting the system's ability to maintain excellent polarization quality even under high thermal loads at peak performance (Fig. \ref{DOP-1GHz}).
Spectral analysis of the amplified signals across a narrow spectral range for different output powers is presented in Fig. \ref{Spectrum-Narrow-1GHz}, showing a consistent 3 dB spectral bandwidth of around  $\approx$ 
 0.51 nm, even at maximum output power.
 In addition, Fig. \ref{Spectrum-Wide-1GHz} displays the optical spectra of the amplified signals across a broad spectral range for different power levels. The spectra reveal $\approx$ 26.6 dB ASE background, but no evidence of SRS, which suggests the absence of significant Raman effects, indicating efficient direct amplification of picosecond pulses.
 Fig. \ref{M2evolution-1GHz} and \ref{M2curve-1GHz} provide a characterization of the beam quality for the 1 GHz laser system based on \(M^2\) factor measurement. The evolution of the beam quality with increasing the output power, along with the corresponding beam profiles, is depicted in Fig.\ref{M2evolution-1GHz}. This measurement shows that the intensity beam profile remains highly symmetric. At maximum power, the beam quality measured as \(M^2X/M^2Y=1.36/1.36\), indicating no sign of mode instability or degradation in beam shape; see Fig. \ref{M2curve-1GHz} (insets: near-field beam profile of the beam at 645 W average power).

\section*{Discussion}
In this report, we demonstrated the capability of all-glass sT-DCF for direct and efficient amplification of short-pulsed lasers with low input average power, maintaining a high degree of polarization and beam quality in a wide range of RR. The sT-DCFs featured ultra-large mode area, enabling amplification of a signal with a few tens mW average power up to 645 W average power or 2 MW peak power in a single-amplifier stage, preserving excellent single-beam propagation and linear polarization state simultaneously mitigating the SRS signal growth. The measurement results for amplifying three different RR laser systems, 1 MHz, 20 MHz, and 1 GHz are summarised in the table \ref{Tabeofcomparison}.  In the amplification of 20 ps pulses with 1 MHz RR, 155 W average output power was achieved with a slope efficiency of 59\%. This corresponds to a remarkably high peak power of 2 MW. The maximum power level was limited by the ASE and onset of DOP degradation, which we attributed mainly to the PMI effect appearing in the weakly birefringent fiber. By increasing the repetition rate of the GS seed laser from 1 MHz to 20 MHz, the same sT-DCF amplifier demonstrated higher efficiency and better maintenance of DOP, 76.6\%, and 88.3\%, respectively. The 3 dB spectral bandwidth was broader at 1 MHz compared to 20 MHz, measuring 0.67 nm and 0.43 nm, respectively. This difference was attributed to the higher levels of nonlinearity experienced during amplification at 1 MHz, where higher intensity per pulse leads to stronger nonlinear effects like self-phase modulation, resulting in greater spectral broadening compared to the repetition rate of 20 MHz. Moreover, the signal-to-ASE ratio at the maximum achieved power was higher at 20 MHz RR (24.3 dB) compared to 1 MHz RR (16.4 dB). This is attributed to more efficient population inversion depletion at higher repetition rates, which leads to less ASE build-up during amplification and thus results in a cleaner, higher-quality output signal. Further power scaling for the amplification of pulses at 20 MHz was limited by pump power. 
For amplification of high RR pulses, 20 ps pulses at 1 GHz RR were amplified up to 645 W, the highest average power achieved for narrow linewidth pulses using sT-DCFs to date. The amplification showed a high slope efficiency of 78.6\%. Further power scaling for the amplification of pulses at 1 GHz was limited by the available pump power.
The sT-DCF amplifier indicated a negligible Raman signal with a level 45.7 dB lower than the main signal peak, which appeared only in the amplification of a 20 MHz pulsed signal. Regarding the beam quality analysis, both the 20 MHz and 1 GHz lasers exhibited beam quality factors of \(M^2 <1.36 \), indicating nearly diffraction-limited beam quality. In contrast, the 1 MHz laser showed a noticeably higher \(M^2\) factor of 2, which we attributed mainly to the nonlinear signal generation in the orthogonal polarization states due to the PMI effect. Although the \(M^2\) factor of 2 typically indicates higher-order mode generation, the beam remained symmetrical, featuring a single intensity maximum without any visible spatial modulation effects. 

\newcolumntype{P}[1]{>{\centering\arraybackslash}p{#1}}
\begin{table}[!t]
  \centering
  \caption{Summary of properties and experimental results of amplification of 1 MHz, 20 MHz, and 1 GHz seed lasers.}
  \begin{tabular}{|P{4.5cm}|P{3.5cm}|P{3.5cm}|P{3.5cm}|}
    \hline
   \textbf{Properties}  & \textbf{1 MHz RR, 50 ps}  & \textbf{20 MHz RR, 50 ps}  &\textbf{1 GHz RR, 20 ps} \\
\hline
\multicolumn{1}{|c|}{Taper Length} & \multicolumn{3}{c|}{6.7 m}    \\
\hline
\multicolumn{1}{|c|}{Diameters variation of

core/first clad/second clad} & \multicolumn{3}{c|}{ 8.3 / 75 / 90 \(\mu \)m  

          to

 90 / 814 / 977 \(\mu \)m}

\\
\hline
\multicolumn{1}{|c|}{NA of core/first clad/second clad} & \multicolumn{3}{c|}{ 0.08/ 0.27/ 0.48 } 
\\
\hline
\multicolumn{1}{|c|}{ Core Absorption} & \multicolumn{3}{c|}{800 dB/m} 
\\

\hline

Average output power  & 155 W

(limited by ASE and DOP degradation) & 625 W

(limited by pump power) & 645 W 

(limited by pump power)\\

\hline

Peak power & 2 MW & 625 kW & 32 kW\\
\hline
Slope efficiency & 59\% & 76.6\% & 78.6\% \\
\hline
Amplification gain & 44.91 dB & 39.82 dB & 38.09 dB \\
\hline
Average of the DOP 

(at the maximum output power) & 70\% & 88.3\% & 87.6\% \\
\hline
Signal-to-Raman ratio

(at the maximum output power) &

No raman signal & 45.7 dB & No raman signal \\
\hline
Signal-to-ASE ratio

(at the maximum output power) &  

16.4 dB & 24.3 dB & 26.6 dB \\
\hline
3 dB spectral bandwidth

(at the maximum output power) & 0.67 nm & 0.43 nm & 0.51 nm \\
\hline
Beam Quality factor

(at the highest output power)&
\(M^2X/M^2Y=2.0/2.0\) &
\(M^2X/M^2Y=1.28/1.34\) &
\(M^2X/M^2Y=1.36/1.36\)\\
\hline
  \end{tabular}
  \newline\newline
 \label{Tabeofcomparison}
\end{table}

\section*{Conclusion}
In conclusion, this study demonstrated a compact, high-power monolithic all-glass spun tapered double-clad fiber amplifier for direct amplification of narrow linewidth picosecond pulses from a few tens of mW to several hundreds of Watts in a single-stage amplification, covering a wide range of repetition rates. The amplifier with a bulk elements-free configuration neglected the alignment failure of the nearly kW pump signal, leading to the better performance of the high-power laser system. The successful amplification of three different repetition rate pulsed signals in the all-glass sT-DCF-based amplifier was demonstrated: 1 MHz (50 ps) over 2 MW peak power, 20 MHz (50 ps) up to 625 W average power, and 1 GHz (20 ps) up to 645 W average power, exhibiting excellent spectral, spatial, and polarization characteristics. The system maintained a high degree of linear polarisation at maximum power levels, achieving  70\%, 88.3\%, and 87.6\% for the 1MHz, 20 MHz, and 1 GHz lasers, respectively. High peak power pulses exhibited beam quality of \(M^2 \sim 2.0 \) at the highest power level with a slope efficiency of exceeding 59\%, while, high average power pulses exhibited near diffraction limited beam quality, \(M^2 <1.36 \) at the highest average output power levels with a slope efficiency of more than 76\%. These results represent significant advancements in narrow linewidth short pulsed laser amplifiers utilizing all-glass ultra-large mode area fiber with remarkable spatial and temporal characteristics and a high degree of polarization, which are suitable for further power/energy scaling of short pulses in a wide range of repetition rates.

\bibliography{main}

\section*{Acknowledgements}

The authors thank Evgenii Motorin, Andrey Chumachenko, Konstantin Miroshnichenko, and Andrei Gurovich for their valuable contributions to this work. The authors would like to thank Prof. G. Genty for the valuable discussion about the PMI effect.

\section*{Funding.}
Part of this work has been supported by the European Commission Horizon2020 program (PULSE project- grant agreement Nr. 824996) and the Flagship for Photonics Research and Innovation (PREIN). 

\section*{Author contributions}
H.F. and E.A and R.G. and V.F. conceived the experiment(s),  H.F. and E.A. conducted the experiment(s), H.F and E.A. analyzed the results. A.G. and E.G. made the taper, A.Y. and E.R. made the GHz seed laser, H.F. wrote the manuscript.  All authors reviewed the manuscript. 

\section*{Competing interests}
The authors declare no conflicts of interest.

\section*{Data availability statement.}
The datasets used and/or analysed during the current study available from the corresponding author on reasonable request.

\section*{Additional information}
\textbf{Correspondence} and requests for materials should be addressed to H.F.

\noindent\textbf{Reprints and permissions information} is available at \href{https://www.nature.com/reprints}{\textcolor{blue}{www.nature.com/reprints}}.

\noindent\textbf{Publisher’s note}Publisher’s note Springer Nature remains neutral with regard to jurisdictional claims in published maps and 
institutional affiliations.

\end{document}